\documentclass[12pt,preprint]{aastex}

\tighten

\def\Sec{${}^{\prime\prime}$\llap{.}}
\def\Min{${}^{\prime}$\llap{.}}

\slugcomment{To appear in {\it The Astronomical Journal}}

\shorttitle{Variable stars in Phoenix}
\shortauthors{Gallart et al.}

\begin{document}


\title{The variable star population in Phoenix: coexistence of Anomalous and short-period Classical Cepheids, and detection of RR Lyrae variables.}


\author{C. Gallart\altaffilmark{1} and A. Aparicio}
\affil{Instituto de Astrof\'\i sica de Canarias. 38200 La
Laguna. Tenerife, Canary Islands. Spain.}
\email{carme@iac.es, aaj@iac.es}

\author{W.L. Freedman}
\affil{Observatories of the Carnegie Institution of Washington. 813 Santa Barbara St. Pasadena CA 91101, USA}
\email{wendy@ociw.edu}
\author{B.F. Madore}
\affil{NASA Extragalactic Database, Infrared Processing and Analysis Center, California Institute of Technology, MS 100-22, Pasadena, CA 91125, USA}
\email{barry@ipac.caltech.edu}
\author{D. Mart\'\i nez-Delgado}
\affil{Max Planck Institute for Astronomy, K\"onigstuhl 17, D69117 Heidelberg, Germany}
\email{ddelgado@mpia-hd.mpg.de}

\and

\author{P.B. Stetson}
\affil{Herzberg Institute of Astrophysics, National Research Council, Victoria, BC, Canada~V9E~2E7}
\email{Peter.Stetson@nrc.gc.ca}




\altaffiltext{1}{Ramon y Cajal Fellow. Instituto de Astrof\'\i sica de 
Canarias. 38200 La Laguna. Tenerife. Spain.}


\begin{abstract}

We present the results of a search for variable stars in the Local
Group dwarf galaxy Phoenix. Nineteen Cepheids, six candidate
long-period variables, one candidate eclipsing binary and a large
number of candidate RR Lyrae stars have been identified. Periods and
light curves have been obtained for all the Cepheid variables. Their
distribution in the period--luminosity diagram reveals that both
Anomalous Cepheids (AC) and short-period Classical Cepheids
\hbox{(s-pCC)} are found in our sample. This is the first time that
both types of variable star are identified in the same system even
though they likely coexist, but have gone unnoticed so far, in other
low-metallicity galaxies like Leo~A and Sextans~A. We argue that the
conditions for the existence of both types of variable star in the
same galaxy are a low metallicity at all ages, and the presence of
both young and intermediate-age (or old, depending on the nature of
AC) stars. The RR Lyrae candidates trace, together with the well
developed horizontal branch, the existence of an important old
population in Phoenix. The different spatial distributions of s-pCC,
AC and RR Lyrae variables in the Phoenix field are consistent with the
stellar population gradients found in Phoenix, in the sense that the
younger population is concentrated in the central part of the
galaxy. The gradients in the distribution of the young population
within the central part of Phoenix, which seem to indicate a
propagation of the recent star formation, are also reflected in the
spatial distribution of the s-pCC.

\end{abstract}



\section{Introduction} \label{intro}

Variable-star studies are experiencing a spectacular resurgence thanks
to the microlensing projects which, as a by-product, are supplying
huge samples of variable stars, mainly in the LMC, SMC and the
Galactic Bulge. On another hand, HST and many ground-based sites with
excellent seeing are allowing us to probe deep into the stellar
populations of Local Group dwarf irregular (dIrr) galaxies, thus
providing information on variable stars only observable before in the
dwarf-galaxy satellites of the Milky Way. These enlarged samples of
variable stars of different types provide the necessary information to
deepen our understanding of the characteristics of each type, the
relationships between them, and the physical mechanisms involved in
their light variation, which yield important tests of stellar
evolution models. At the same time, the gathering of data in different
galaxy environments makes it possible to relate the types of variable
stars to the stellar populations in their host galaxies.

These studies have provided, in particular, relatively large samples
of short-period Classical Cepheids (s-pCC) in a number of dIrr
galaxies: the SMC (Bauer et al.\ 1999; Udalski et al.\ 1999), IC1613
(Dolphin et al.\ 2001), Leo~A (Dolphin et al.\ 2002) and Sextans~A
(Dolphin et al.\ 2003). As pointed out by Gallart et al.\ (1999) and
Dolphin et al.\ (2002), short-period (P~$\simeq$~1 day) variables,
with Cepheid-like light curves and luminosities about 1 magnitude
above the horizontal branch can be produced by
young\footnote{Throughout the paper, we will consider as young
populations those stars younger than 1 Gyr, while intermediate-age
populations will be defined broadly as those having ages of 1--10 Gyr}
stars in the phase of core He burning which, in the case of
low-metallicity stars, experience blue loops extended enough to the
blue to cross the instability strip. The so-called Anomalous Cepheids
(AC) and s-pCC lie in similar positions in both the color--magnitude
and the period--luminosity (PL) diagrams, and they are found in
basically every dwarf Spheroidal (dSph) galaxy that has been surveyed
for them, as well as in one globular cluster (Pritzl et al.\ 2002; and
references therein). The term ``Anomalous Cepheid'' was first
introduced by Norris \& Zinn (1975) because they are more luminous at
a given period than the Population II Cepheids found in globular
clusters. AC are $\simeq\,$0.5--2 magnitudes brighter than RR Lyrae
stars, and their periods range from 0.3 to 2 days. Concerning their
evolutionary status, there is general agreement that they are
metal-poor stars with mass $\simeq 1.5 M_\odot$, occupying the
instability strip during their horizontal-branch phase of evolution
(Demarque \& Hirshfeld 1975; Hirshfeld 1980; Bono et al.\ 1997). They
may be either i)~evolved, single, intermediate-age ($\le 5$ Gyr)
stars, or ii)~the evolved products of mass transfer in old ($\simeq
12$ Gyr) binary systems.  The predicted behaviour in both cases is
very similar (Bono et al.\ 1997), and it has therefore been considered
impossible to distinguish between the two possible origins of AC from
their location in the PL and color-magnitude diagrams alone (Nemec,
Nemec \& Lutz 1994 and Dolphin et al.\ 2002).


A reanalysis of all the
data accumulated to date on AC by Bono et al.\ (1997) and Pritzl et
al. (2002) provided a slightly different locus for the AC in the PL
diagram that, as we will show in the present paper, may allow one to
discriminate these two types of variable star. Since they
respectively trace stellar populations of different ages and of a very
specific metallicity range, this distinction is relevant to the use of
variable stars as bright tracers of stellar populations of different
ages (in this case, AC as tracers of a fainter intermediate-age or old
population), and as anchors of the age--metallicity relation of
individual dwarf galaxies.

Prior to this study, AC and s-pCC have not been identified in the same
galaxy.  As mentioned above, AC are routinely found in dSph galaxies
while s-pCC have been found in dIrr galaxies.  Since both types of
variable star have very similar characteristics, it may be that the
classification of some of them has been prejudiced by the type of
galaxy in which they were found. A suitable type of object in which to
search for {\it both\/} types of variable are the so-called
transition-type dSph/dIrr galaxies (Mateo 1998).  The closest such
system is Phoenix. We had started a search for variable stars in this
galaxy, and along the way we realized that it was, in fact, an ideal
system to address this problem.

Phoenix is a low-mass ($M_{tot}=3.3\times10^6 M_{\odot}$),
low-metallicity ([Fe/H]$\,$=$\,$--1.4) system at a distance of about 400 Kpc
[$(m-M)_\circ=23.0 \pm 0.1$] from the Milky Way (Mateo 1998; Mart\'\i
nez-Delgado, Gallart \& Aparicio 1999, MGA99 hereinafter; Held,
Saviane \& Momany 1999; and references therein).  The fact that it has
a small young stellar population (up to about 100 Myr old), embedded in a
substantial population of old and intermediate-age stars (MGA99; Held
et al.\ 1999) and little or no gas (St-Germain et al.\ 1999; Gallart et
al. 2001) has motivated its classification as a transition type
dSph/dIrr galaxy. No studies of variable stars in Phoenix have been
published to date, although some suspected variable stars were reported
by van~de~Rydt, Demers \& Kunkel (1991) and MGA99. Both its
low metallicity and the existence of stars of all ages made it a good
candidate to search for both AC and s-pCC. In this paper we report such
a discovery. We also discuss the possible existence of both types of variable star in
other dwarf galaxies and the information that they may provide on the
age--metallicity relations of their host galaxies. We also report, for the
first time, the discovery of a large number of RR Lyrae variable star
candidates in Phoenix which trace, together with the well developed
horizontal branch, the existence and distribution of an old population in the 
galaxy.

\section{Observations and data reduction} \label{obs}

Phoenix was observed in 1997, 2000, and 2001 during four approximately
week-long runs using the 2.5m duPont telescope at Las Campanas
Observatory.  In addition, we included one epoch obtained by ourselves
with FORS1 at the VLT and several epochs of ESO archival
data obtained with either the NTT (with EMMI or SUSI) or the VLT.
Since our initial goal was the study of the Cepheid population in
Phoenix, we typically performed observations once per night in two
bands ($V$ and either $B$ or $I$) during each of the Las Campanas runs, except
on the first run, when several images per night were obtained for
the stellar-populations study published by MGA99. All the
observations available to us are listed in Table~\ref{obslog}. Only
images with seeing better than 1\Sec5 were used for the present
study.

The images were processed in the standard way using IRAF to apply bias
and flat-field frames obtained on the corresponding
nights. Profile-fitting photometry was obtained using the
DAOPHOT/ALLSTAR/ALLFRAME suite of programs \citep{stetson87,
stetson94}. A master frame for Phoenix was created from the median of
60 $B$, $V$ and $I$ LCO images with FWHM $\le$ 4 pix (i.e., seeing
$\le$ 1\arcsec). Geometric transformations between frames, including
cubic distortions in addition to translations and scale changes, were
calculated using DAOMASTER \citep{stetson90}. MONTAGE2 was used to
create the medianed image, which was then searched for stars.  A total
of $\simeq$ 14,700 stars were found and input to ALLFRAME, to be
photometered on a total of 19 $B$, 48 $V$, and 45 $I$ good-seeing
($\le$ 1\Sec5) images. Mean $B$, $V$, and $I$ magnitudes at each epoch
were obtained for each star in the corresponding set of images in each
filter, again using DAOMASTER. Aperture corrections were obtained from
a large number of isolated stars in one image for each of $B$, $V$,
and $I$. These are estimated to be accurate to $\pm$ 0.003 mag in $B$,
$\pm$0.008 mag in $V$, and $\pm$0.006 mag in $I$. Photometric
transformations to the Johnson-Cousins standard system of Landolt
(1992) were obtained using standard-star measurements in two of the
observing runs.

In February 1997, we obtained approximately 140 measurements of
Landolt standard stars in $V$ and $\simeq$ 120 in $I$, providing a
good sampling at all airmasses up to X$\simeq$2.0. They were used to
obtain atmospheric and instrumental coefficients for transforming our
instrumental magnitudes to the Johnson $V$ and Cousins $I$ standard
photometric systems. These same stars were used to obtain the
instrumental coefficients published in MGA99, but a later reanalysis
of the same data showed that an error had been made in the sign of one
of the color terms. The corrected equations used to transform our
instrumental magnitudes into the Johnson-Cousins system are:

$$V-v = 24.766 - 0.029 (V-I)~~~  (1) $$
$$I-i = 24.443 + 0.032 (V-I)~~~  (2) $$

The photometric conditions during the campaign were stable and produced
very small zero-point errors for the photometric transformations: $\pm$
0.006 mag in $V$ and $\pm$ 0.009 mag in $I$. The standard errors in the
extinction were always smaller than 0.012 in both $V$ and $I$.  Taking all
the uncertainties into account, we estimate a total error in the
photometry zero point of 0.02 mag in both $V$ and $I$.

The most important difference between these equations and those listed
by MGA99 is in the sign of the color term of equation (2) (we also
find a different numerical value for the zero point, but this is
mostly compensated by a corresponding change in the calculated
extinction coefficients). With this new color term, the discrepancy in
the color of the RGB tip between MGA99 and both van~de~Rydt et al.\
(1991) and Held et al.\ (1999) --- which was acknowledged by MGA99 ---
has disappeared.

The $B$ transformations were determined from 10 observations of 3
Landolt (1992) standard stars in the December 2001 observing run.
The corresponding transformation equation is:
$$B-b = 24.547 + 0.043 (B-V)~~~  (3) $$

\noindent The total uncertainty in the $B$ zero-point (extinction, photometric
transformation, and aperture correction) is estimated to be $\pm 0.02$.

\begin{figure}
\plotone{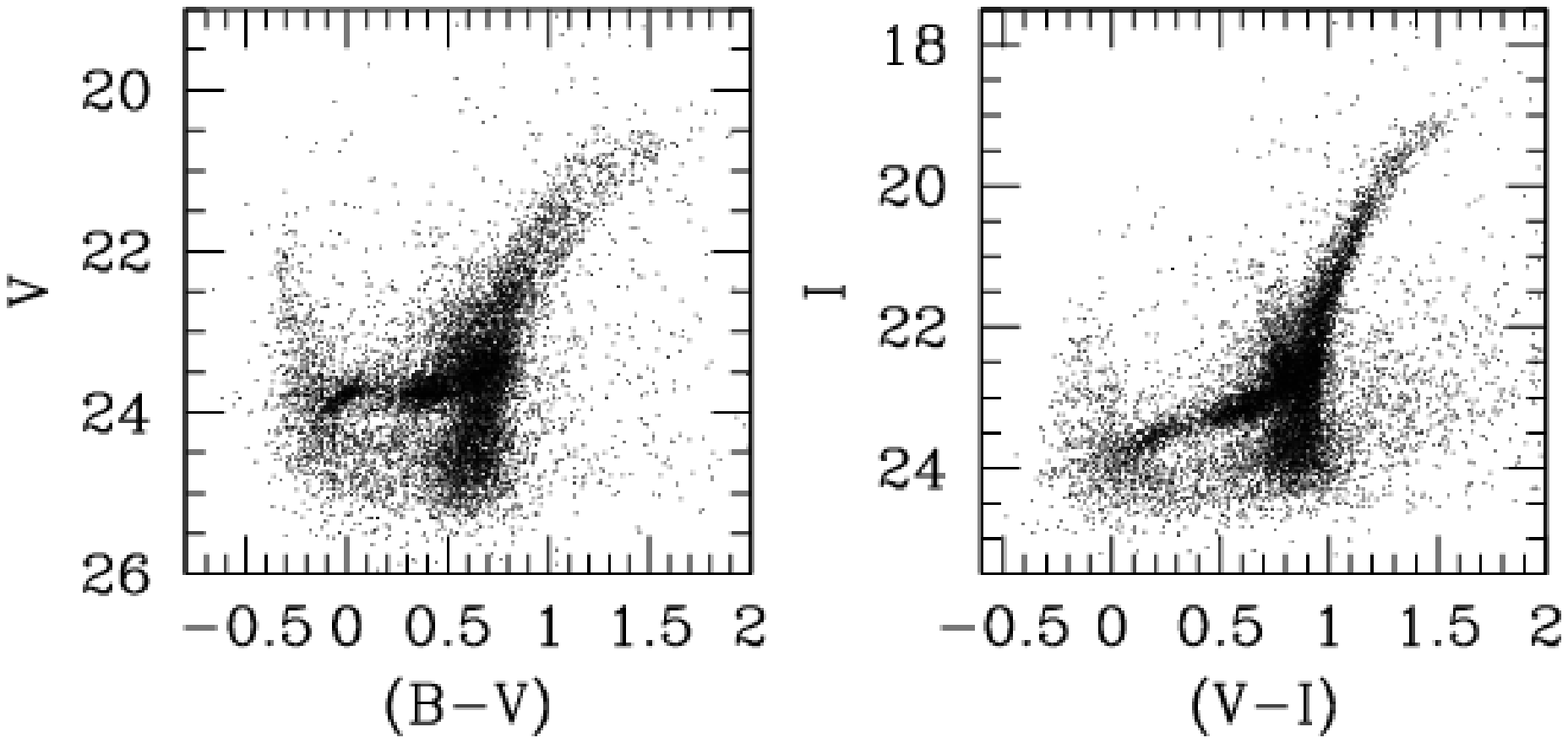}
\caption{[(B-V),V] and [(V-I),I] CMDs of Phoenix.}
\label{cmds}
\end{figure}

Figure~\ref{cmds} shows the [({\it B--V\/}),$V$] and [({\it
V--I\/}),$I$] CMDs obtained by combining all the data. Note that a
blue-extended, well populated horizontal branch (HB), which was hinted
in both MGA99 and Held et al. (1999), shows up beautifully in both
CMDs. An RR Lyrae gap also clearly appears in both diagrams.

\section{Selection of candidate variable stars}

\begin{figure}
\plotone{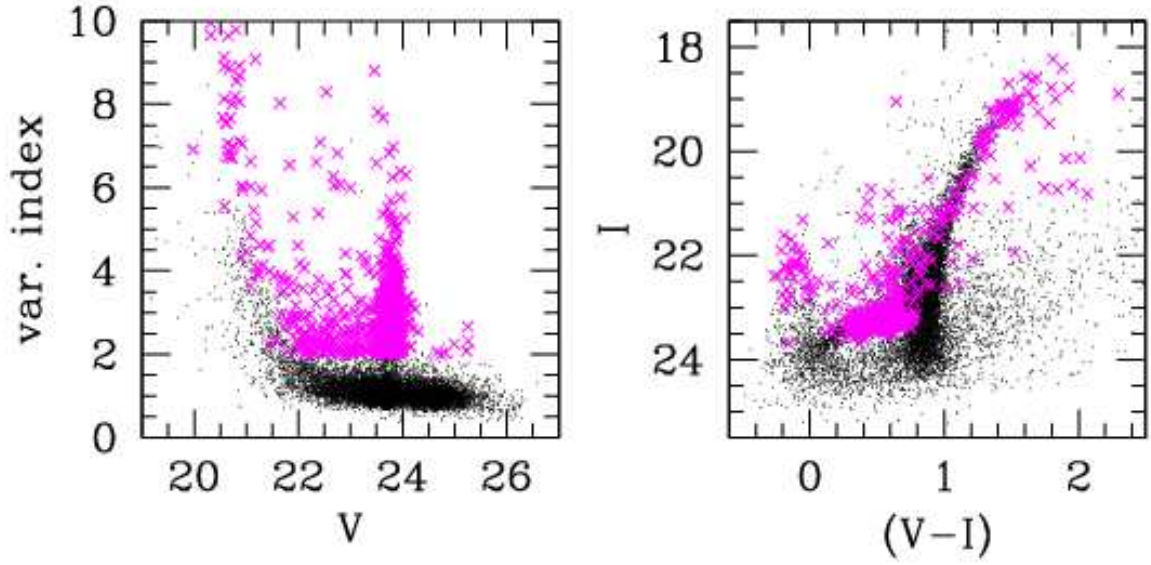}
\caption{Left: First-pass candidate variable stars (crosses). 
Right: [({\it V--I\/}), $I$] CMD of Phoenix, with the first-pass candidate
variable stars highlighted. A substantial number of spurious
detections appear in this sample.}
\label{candi1}
\end{figure}

We made a first selection of candidate variable stars based on both
the variability index $var$, defined as the ratio of the external
($\sigma_{ext}$) to the internal ($\sigma_{int}$) standard
error\footnote{The {\it internal\/} standard error of the magnitude
measured for a single star in a single image, $\sigma_{int}$, is
estimated from the known readout noise of the detector and the Poisson
photon noise of star and sky as scaled by the known gain of the
detector, and also from the pixel-to-pixel agreement between the
observed stellar profile and the shifted and scaled point-spread
function for the image.  The {\it external\/} standard error per
observation for a given star, $\sigma_{ext}$, is derived from the
frame-to-frame agreement between its individual magnitude
measurements.  A star that is actually varying in luminosity and that
has been well measured at most or all epochs should have $\hbox{\it
var\/} \equiv \sigma_{ext}/\left<\sigma_{int}\right>\gg1$, where
$\left<\sigma_{int}\right>$ is a suitable estimate of the ``typical''
internal error for the star.  This index cannot be justified with full
mathematical rigor; nevertheless, it is apparent that that minority of
stars having stand-out values of {\it var\/} are the most likely to be
genuine variables.}, and on the external standard error of one
measurement ($\sigma_{ext}$) itself, as calculated by DAOMASTER from
the individual magnitudes of the 48 $V$ images we ultimately used.
The stars with $var \ge 2.0$ and $\sigma_{ext} \ge 0.05$ were flagged
as candidate variable stars. The second requirement was set in order
to discard many bright stars whose large variability index originated
in a very small internal standard error (which is itself a derived
statistic subject to observational noise and round-off error), and
thus were not {\it bona fide\/} candidate variable stars.  Also, the
area corresponding to the brightest star in the field was masked,
since the star image was generally saturated.  These requirements
produced a list with about 800 variable-star candidates.  These
candidates were then individually examined using the image display to
discard problematic stars: any candidate i)~in the wings of a
companion star or galaxy much brighter than itself, or ii)~with a
companion star that ALLFRAME failed to separate, or iii)~with poorly
subtracted neighbors. In this way, about 25\% of the candidates were
discarded.  Even though some of these criteria may discard {\it bona
fide\/} variable stars, they would be stars with poorly determined
magnitudes or with magnitude errors likely to be sensitive to the
seeing, for which accurate periods would be difficult to determine.
This left a list of $\simeq$ 590 stars; these remaining candidates are
represented in Figure~\ref{candi1}. Note that the fact that there are
many candidates among the brightest blue and red stars, where they are
not {\it a priori} expected, indicates that some of them will still be
spurious\footnote{These spurious candidates still originate in stars
that have very small internal errors $\sigma_{int}$ resulting in large
values of the $var$ index, which indicates that the condition
$\sigma_{ext} \ge 0.05$ has not been stringent enough. We preferred,
however, to retain these stars and search for variability on them,
rather than risk discarding possible {\it bona fide\/} low-amplitude
variables.}.

Subsequently, these 590 candidate variable stars were divided into two
groups: those above the HB, {\it i.e.}, brighter than $V\simeq 23.2$
which would be Cepheid, eclipsing or long-period variable candidates,
and those fainter, which are RR~Lyrae variable-star candidates. No
candidates were discarded {\it a priori} based on their color, since
we are interested in all kinds of variable stars, and not only
pulsating variable stars in the instability strip. In addition, the
{\it mean} magnitudes computed at this point were just straight
averages of all data points (weighted by the photometric errors), and
not properly intensity-averaged, phase-weighted magnitudes. The colors
computed from them, therefore, may be affected by a different sampling
of the the light curve in the two bands, and therefore substantially
different from actual mean colors\footnote{Note however, that the
color terms in the photometric transformations are small (0.03--0.04
mag), so that even if the colors are wrong by 1.0 magnitudes, the
individual calibrated magnitudes would still only be off by 0.03--0.04
mag due to the incorrect inferred colors.  In addition, one iteration
in the calculation of the individual calibrated magnitudes was
performed: in a first pass, the straight average $B$, $V$ and $I$
magnitudes were used to calculate the colors of each star to be used
in transforming the instrumental magnitudes to standard magnitudes
using eqs. (1--3). Subsequently, intensity-averaged, phase-weighted
magnitudes were calculated for each star, and used in a refined
transformation of the individual-epoch magnitudes.}. This, in fact,
appears to be the case for two of the {\it bona fide\/} pulsating
variables discovered, which had very blue inferred ({\it B--V\/})
colors (see below).

\subsection{The bright candidates} \label{bright}

The selection described above produced a list of 202 {\it bright}
(V$\le$23.2) candidates. The light variation of these stars over all
epochs was examined graphically and compared with the magnitude
scatter of non-variable stars of similar magnitude. In the process, a
large fraction of candidates was discarded because their light
variations were similar to the average magnitude dispersion at the
corresponding magnitude. From the characteristics of the light
variation, 47 candidates seemed periodic, 17 were candidate eclipsing
binaries, and 9 were candidate long-period variables. All eclipsing
binary candidates but one were discarded after examination of the star
image in the frame that corresponded to minimum light, in most cases
because it was on a bad column of the detector.  Among the long-period
variable candidates, however, 6 are likely legitimate variable stars.

The light variations of all the candidate pulsating ``bright'' variables
(likely s-pCC or AC) were searched for periodicities. We used the
three following methods: i)~the Stellingwerf (1978) phase dispersion
minimization algorithm; ii)~our own modified string-length algorithm, which
considers the data from all filters simultaneously and takes into
account the photometric uncertainty associated with each data point;
and iii)~the routines created by Andy Layden (Layden 1998; Layden \&
Sarajedini 2000), which determine the most likely
period by fitting the photometry of the variable star with 10
templates over a selected range of periods.  For methods i) and iii) we used
only the $V$ data points.  In most cases, the same best period
was found by all three methods. In only a few cases, one of the three
methods produced a period that the other methods didn't find,
and which turned out to be the best upon examination of the light
curves.

We visually inspected the light curves that each of the most likely
periods would produce for each variable, in {\it all three\/}
bands. When more than one likely period was found, the quality of the
derived light curve was used to decide among them. In a few cases,
more than one period produced acceptable light curves. In these cases,
we used the position in the period-luminosity (PL) diagram (see
Section~\ref{nature_cep}) to identify the most
likely periods (see Table~\ref{candidates}).

\begin{figure}
\epsscale{1.0}
\plottwo{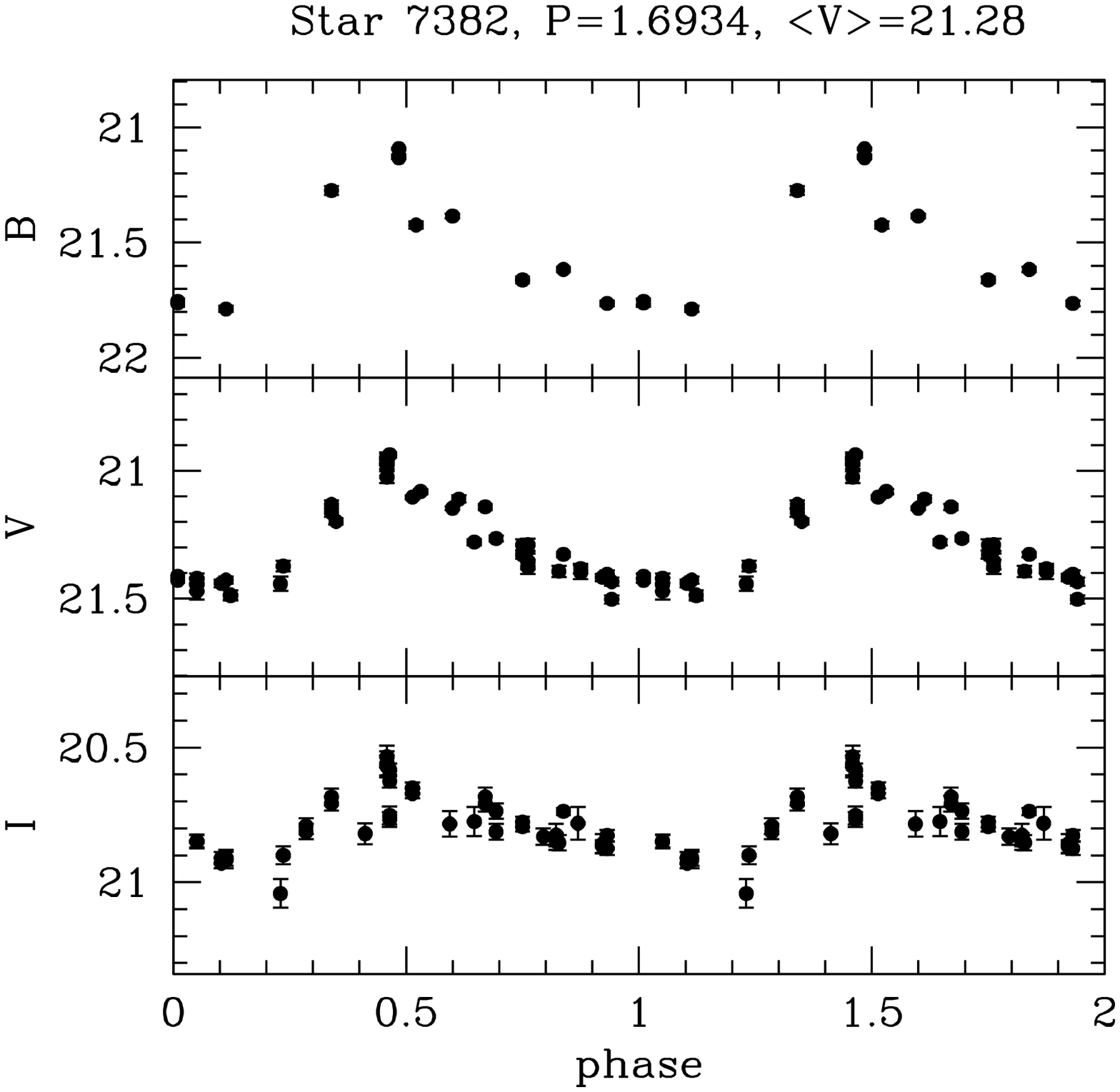}{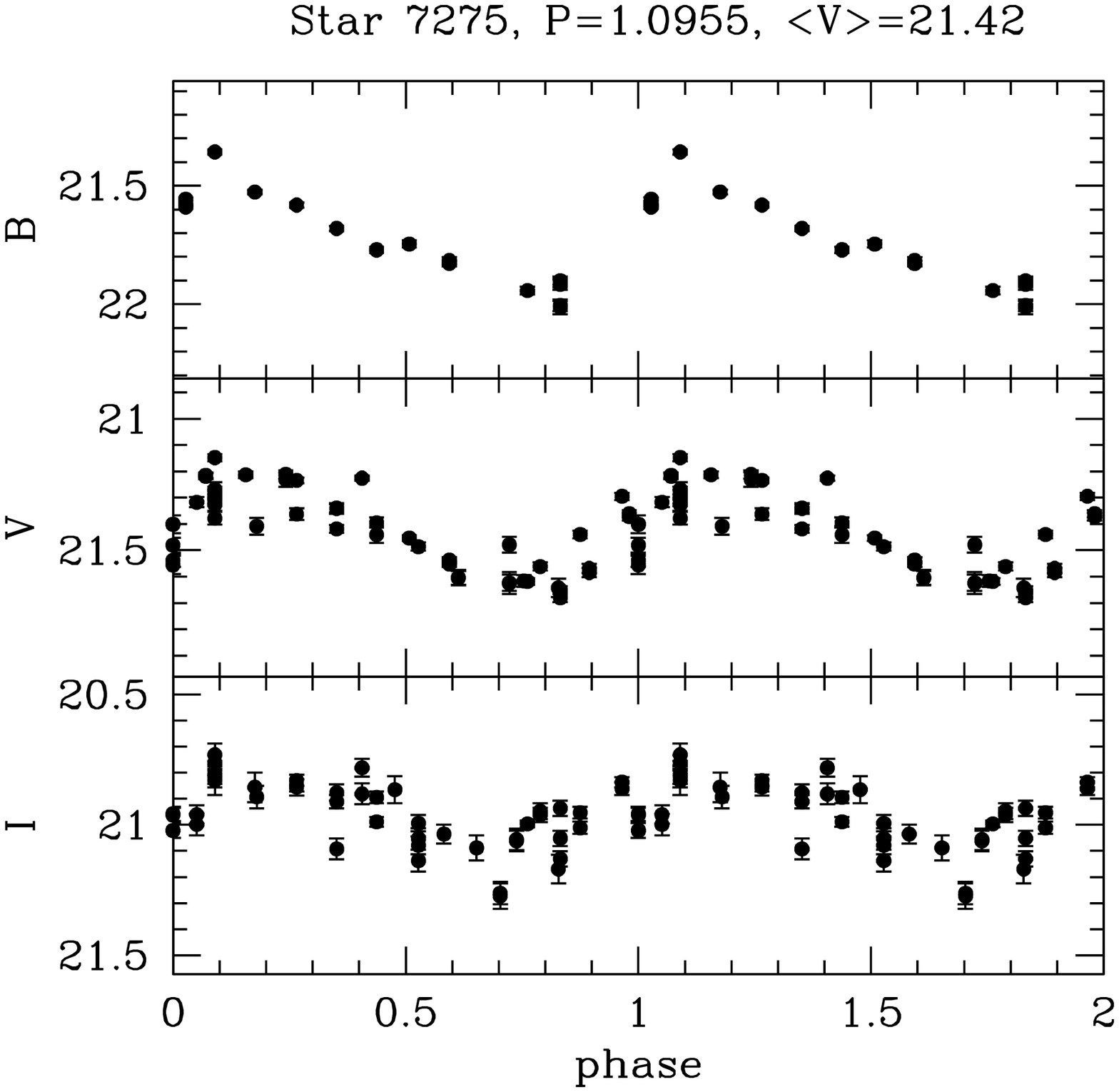}
\caption{a) $B$, $V$ and $I$ light curves for the Phoenix Cepheids. Data are repeated over a second cycle for clarity.}
\label{lccep}
\end{figure}

For 15 candidates, clear periods were found, while 4 have uncertain
periods. The periods are all shorter than 2 days, and about half of
the variables have periods shorter than 1 day. The PL distribution of
the Phoenix variables will be discussed in Section~\ref{pl}, together
with the nature of these variable stars. In Table~\ref{candidates} the
identification of each variable, its period, the phase-weighted
intensity-averaged magnitude (Saha \& Hoessel 1990) in each band, and
its most likely sub-class are listed. The intensity-averaged
magnitudes (not phase-weighted) were also calculated for all these
variables. The differences from the phase-weighted magnitudes are of a
couple of tenths of a magnitude at most. The phase-weighted magnitudes
produce tighter sequences in the CMD and are preferred as a best
representation of the true mean magnitude. The {\it BVI\/} light
curves for these stars are displayed in Figure~\ref{lccep}.

The final list of confirmed pulsating and long-period variables, and
the one candidate eclipsing binary star are listed in
Table~\ref{candidates}.  A finding chart is displayed in
Figure~\ref{mapas}. The $B$, $V$, and $I$ photometry of the variables
at the different epochs is listed in Tables~\ref{bphotometry},
\ref{vphotometry} and ~\ref{iphotometry} respectively.  They are shown
in Figure~\ref{dcmvilim}, together with all the candidate RR Lyrae
variable stars (see Section~\ref{rr}).

\subsection{The RR Lyrae candidates} \label{rr}

The RR Lyrae candidates are close to the photometric limit of our CMD,
and in most cases, their light variations are similar to the
photometric errors expected at the corresponding magnitude level. For
most of them no definite period can be assigned, and an exhaustive
search has not been performed. The great concentration of candidate
variable stars in the HB (see Figure~\ref{candi1}), with almost no
variable stars with redder or bluer colors at similar magnitudes,
however, supports the idea that we are indeed detecting the light
variations of RR Lyrae variables. A few light curves for RR Lyrae for
which clear periods have been found are shown in Figure~\ref{lc_rr}.

\section{Nature of the Cepheid variables in Phoenix: Classical or 
Anomalous?} \label{nature_cep}

In Figure~\ref{pl} the PL diagrams for the Phoenix Cepheids in the $B$
and $V$ bands are represented. A distance modulus $(m-M)_\circ=23.1$
and a reddening $E(\hbox{\it B--V\/})=0.02$ (Held et al.\ 1999) have
been assumed. Because the number of Cepheid variables found in this
work is relatively small (as expected in a low-mass system with little
current star formation), we prefer to use the tip of the RGB as a
robust distance indicator rather than trying to derive a distance from
the variables themselves. Given this value for the true distance
modulus, the pulsating variables located above the horizontal branch
extend from $M_V \simeq -2, M_B \simeq -1.8$ down to $M_V \simeq 0$
and $M_B \simeq 0.3$, virtually merging with the RR Lyrae locus, and
forming therefore a continuum in the instability strip.

To further constrain the nature of these variables from 
their positions in the PL plane, the PL relations
obtained by fitting the s-pCC in the OGLE database, and the PL
relations obtained by Pritzl et al.\ (2002) for Local Group dSph AC
have been overplotted on the Phoenix data in Figure~\ref{pl}.

Because of the change in the PL relation slope at a period of about 2
days (Bauer et al.\ 1999; the effect is also clearly visible in the
OGLE SMC Cepheids), we fit a PL relation for s-pCC in the OGLE SMC
database. Assuming a SMC distance modulus of $(m-M)_\circ=18.66$ and
mean reddening $E(\hbox{\it B--V\/})=0.09$ (Udalski et al.\ 1999), the
PL relations become:

$$M_{B,FM}=-2.66(\pm 0.12){\rm log}(P)-0.92(\pm 0.02)~~~  (4)$$
$$M_{B,FO}=-3.24(\pm 0.06){\rm log}(P)-1.58(\pm 0.01)~~~  (5)$$
$$M_{V,FM}=-3.08(\pm 0.10){\rm log}(P)-1.12(\pm 0.02)~~~  (6)$$
$$M_{V,FO}=-3.31(\pm 0.06){\rm log}(P)-1.78(\pm 0.01)~~~  (7)$$

\noindent As discussed by Dolphin et al (2002), these relations are virtually
identical to the PL relations for AC obtained by Nemec et
al. (1994)\footnote{Nemec et al.\ (1994) give a PL relation for AC
which depends on metallicity. Their relation is closest to the PL
relation for s-pCC at the lowest possible metallicities, while it
departs substantially from the s-pCC under the assumption of a higher
metallicity}. In contrast, the PL relations for AC obtained by Pritzl
et al.\ (2002) are substantially different:

$$M_{B,FM}=-2.62(\pm 0.18){\rm log}(P)-0.40(\pm 0.04)~~~  (8)$$
$$M_{B,FO}=-3.99(\pm 0.27){\rm log}(P)-1.43(\pm 0.09)~~~  (9)$$
$$M_{V,FM}=-2.64(\pm 0.17){\rm log}(P)-0.71(\pm 0.03)~~~  (10)$$
$$M_{V,FO}=-3.74(\pm 0.20){\rm log}(P)-1.61(\pm 0.07)~~~  (11)$$

\noindent The dataset on which Pritzl et al.\ (2002) based their fit to the
AC PL relation is somewhat larger than the dataset available to Nemec
et al.\ (1994), and shows clearly the different slope of the
fundamental mode and first overtone relations, of which there was
already a hint in the work of Nemec et al.\ (who decided,
nevertheless, to keep the two relations parallel in view of the lack
of clear statistical evidence for different slopes). We will use the
Pritzl et al.\ relation for AC, which is also very similar to the Bono
et al.\ (1997) relation, in the remainder of this paper.

In Figure~\ref{pl}, there is some evidence that we find both types of
Cepheid in Phoenix. Down to $M_V \simeq -1$, the variables seem to
follow the SMC s-pCC PL relation quite nicely, while at fainter
magnitudes they follow the AC PL relation much better. In the
remainder of the paper, we will adopt this limit to assign each of the
Phoenix Cepheid variables to one of the two subtypes
(Table~\ref{candidates}, and Figure~\ref{distrixy}). This is the first
time that the two types of Cepheid have been identified in the same
system.  Previously, AC were clearly identified only in dSph galaxies
and globular clusters, while s-pCC were identified in several dIrr
galaxies (see Section~\ref{discussion}). It is fortuitous that the
first identification of both AC and s-pCC has occurred in a galaxy of
the so-called dSph/dIrr transition type (Mateo 1998). We will discuss
in Section~\ref{discussion} that both kinds of variables may have been
observed in other galaxies considered pure dIrr systems, even though
the presence of AC went unnoticed.

\section{Spatial distribution of stellar populations} \label{spatial}

Because AC and s-pCC are representatives of stellar populations of
different ages (basically, older and younger than 1 Gyr, respectively)
we may expect that, if our interpretation of the nature of the
variables in Phoenix is correct, they must show different
distributions consistent with the stellar population gradients that
have already been noted in Phoenix. MGA99 showed that all the young
stars in Phoenix are concentrated in a flattened inner component with
a maximum extent of $\approx 458\,$pc (approximately 4\arcmin)
east-west, while the older stars extend farther out
(diameter$\,>\,900~{\rm pc}\,\simeq\,$7\Min8) as a somewhat flattened
component oriented North--South.  In the inner, young component, a
gradient in the age of the stars is observed, with the youngest stars
($\simeq$ 100 Myr old) predominantly located in its western half, and
likely remnants of gas lost by the galaxy after this last event of
star formation situated 6\arcmin\ southwest from the center of the
galaxy (Gallart et al.\ 2001).

To show that the variable stars are correlated as expected with the
rest of the stellar populations in Phoenix, a few isochrones have been
plotted in the Phoenix CMD (see Figure~\ref{dcmviiso}). Candidate
s-pCC stars are represented as circles and AC as stars (using, as
defined in Section ref~\label{nature_cep}, $M_V=-1$ as the cutoff
magnitude between them). The s-pCC variables are well matched by the
blue loops of low-metallicity ($Z=0.001-0.0004$), 400--600$\,$Myr old
stars. Stars of the same age are also found in the main sequence of
the same CMDs. As calculated by Gallart et al.\ (1999), a star of this
age in the blue-loop phase would have an expected period $P\simeq$1.2
days, which is in the middle of the period range found for these
s-pCC. The ACs are well matched by the "hooks" (Bono 2003) of He-burning 1--2
Gyr old, very metal-poor ($Z=0.0001$) stars.


In Figure~\ref{distrixy} we have plotted the positions of the variable
stars of different types in Phoenix. The s-pCC variable stars (marked
as circles in Figure~\ref{distrixy}), are concentrated in the central
part of the galaxy where the most recent star formation has taken
place. They also show a very flattened distribution, oriented in the
same direction as the young Phoenix component. The AC variable stars
(marked as stars in Figure~\ref{distrixy}) are much more widely
distributed, as is the intermediate-age 
population in Phoenix. The RR Lyrae candidates are even more widely
distributed, and they probably trace the true extension of the 
old population in Phoenix which, as concluded by MGA99, seems to have a
spheroidal distribution extending farther than the younger inner
component.

Because the AC variables in Phoenix are appreciably less widely
distributed on the sky than the RR Lyrae candidates, throughout the
rest of this paper we will tacitly assume that the AC variable stars
in Phoenix have evolved from single intermediate-age stars (age $\sim$
a few Gyr) rather than from mass-exchange binaries in an old ($\sim
10\,$Gyr) population.  We admit that this assertion is not yet
decisively proven; AC of both types of provenance could well exist in
Phoenix or other systems.

Going back in more detail to the spatial distribution of the bright s-pCC
variables, it is interesting that most of them are located preferentially
in the eastern half of the Phoenix central zone, opposite to the locus of
the very youngest population. In principle, one might have expected some
brighter, slightly longer-period variables in the western part of the
galaxy, but these are not found. From Figure~\ref{dcmviiso}, it can be
seen that the population in the western part is as young as 100 Myr. As
discussed by Dolphin et al. (2001), the bright blue loops are less strongly
populated than the fainter ones, and small-number statistics may be the
reason for not finding younger Cepheids. All this is consistent with the
hypothesis of a propagation of the star formation in the central component
of Phoenix as put forward by MGA99: a burst of star formation would have
started $\simeq\,$600 Myr ago in the eastern part, where most of the
variables of that age are found, and propagated westward, where it was
active just $\simeq\,$100 Myr ago.

Finally, in Figure~\ref{perfil}, RGB, RR Lyrae and Cepheid star counts
over several radial bins are displayed. Note the different
distribution as a function of radius of RGB stars and RR Lyrae
candidates: the RGB stars are substantially more centrally concentrated
than the RR Lyrae candidates, whose distribution out to $\simeq$ 70
arcsec from the center of the galaxy is relatively flat (inside this
radius, completeness effects may play a role), and certainly flatter
than that of the RGB stars. This may indicate that the old population
of the galaxy, traced by the RR Lyrae candidates, is more uniformly
distributed than the RGB, which is likely a mixture of old
and intermediate-age populations (in the central 3\arcmin\ of the galaxy
at least). A detailed discussion of the radial distribution of stellar
populations of different ages in Phoenix, using HST data, will be
presented by Hidalgo, Aparicio \& Mart\'\i nez-Delgado 2003, in prep).

\section{Discussion: Anomalous {\it vs.} short period Classical Cepheids: 
why are they rarely found in the same system?} \label{discussion}

From a stellar-evolution point of view, both AC and s-pCC are
central He-burners, the difference between them being the fact that
the AC started the core He burning under degenerate conditions (Bono
et al.\ 1997), whereas s-pCC are stars massive enough to have ignited
He in the core under non-degenerate conditions. The former are,
therefore, older than the latter, as discussed in the Introduction.
As was also discussed and demonstrated by Dolphin et al.\ (2002), the
reason for the existence of large numbers of s-pCC in low-metallicity
systems is that the blue loops extend farther to the blue for stars of
lower metal abundance and, as a consequence, the shorter but more
populated blue loops of relatively low-mass, old stars may enter the
instability strip. Because the "hooks'' of intermediate-age He-burning
stars also reach the instability strip only in the case of low
metallicity (Bono 1997), both AC and s-pCC can exist only in
low-metallicity systems, and the existence of one or the other depends
on the stars of the appropriate age being present in the galaxy. Both
types of variables coexist in Phoenix probably thanks to its very low
metallicity and the presence of stellar populations of all ages.


In some of the dSph satellites of the Milky Way, the relatively
short-period variable stars situated above the horizontal branch must
necessarily be AC, since no other stars young enough to produce blue
loops are observed in them. In others --- like Leo I --- there is a
low-metallicity population young enough to potentially provide s-pCC
(Gallart et al.\ 1999). On the other hand, no AC have been identified
so far in dIrr galaxies. Part of the reason may be that dIrr are too
distant for stars of such low luminosity to be detected. On the basis
of limiting magnitude they could have been detected, however, in the
SMC and LMC, and in IC1613, Leo~A, and Sextans~A (specifically, in the
recent studies by Dolphin et al.\ 2001, 2002, 2003). In the SMC OGLE
database, for example, a comparatively small number of objects fall
along the AC PL relation obtained by Pritzl et al.\ (2002; see also
Smith et al. 1992). The reason for the lack of clear AC sequences in
the LMC and SMC may be that their metallicities are too high.


In the case of Leo~A, Dolphin et al.\ (2002) concluded that all their
variables are compatible with being s-pCC and that, in any case, these
could not be distinguished from AC on the basis of their position in
the PL plane.  (Note that they were considering the Nemec et al.\ PL
relation for AC of very low metallicity, which is basically identical
to that of s-pCC.) Since Leo~A is supposed to be a very
low-metallicity galaxy like Phoenix, and does contain an old
population as evidenced by the presence of a number of RR Lyrae stars
(Dolphin et al.\ 2002), one would expect that AC could also exist in
it. In the following we will show that, in fact, the distribution of
Cepheids in the Leo~A PL diagram is better fit if both AC and s-pCC PL
relations are assumed to be present, as opposed to only s-pCC as
assumed by Dolphin et al.\ (2002). In Figure~\ref{leoa_sex}, Figure 8
of Dolphin et al.\ (2002) is reproduced, but the Pritzl et al.\ (2002)
AC PL relation has been superimposed. Note that a number of stars
situated below the PL relations of both fundamental-mode and
first-overtone s-pCC are nicely fit by the AC PL relations. We think,
therefore, that both types of variables are present in Leo~A as well
as in Phoenix. A similar situation occurs in Sextans~A (see
Figure~\ref{leoa_sex}), even though very few variable stars have been
detected in this galaxy below $M_V=-1$, and thus the main magnitude
range occupied by AC is missing.

We conclude, therefore, that the only requirements for the coexistence
of s-pCC and AC in a given galaxy is that its metallicity be low
enough ($Z\simeq 0.0004$) and that there have been star-formation
activity at all ages. The two populations of variables can be
distinguished in the PL plane and their presence or lack thereof can
provide important hints about the age--metallicity relation of the
host galaxy.

According to the metallicity--absolute magnitude relation observed for
dwarf galaxies, which implies that low-luminosity systems
are also metal-poor, we expect to find both AC and s-pCC in the
smallest dIrr galaxies, e.g. NGC6822 (see Clementini et al.\ 2003),
IC1613, Leo~A, Sextans~A, Sextans~B, Pegasus, LGS3, and Antlia.

Finally, Mateo, Fischer \& Krzeminski (1995) examined the specific
frequency $S$ of ACs (number of AC per $10^5 L_{V,\odot}$) in dSph
galaxies and found a good correlation between $S$ and galaxy
magnitude. Pritzl at al. (2003) updated this relation with all the
currently available data. It is interesting to check whether galaxy
types other than dSph follow the same relation.  In the case of
Phoenix, with 11 AC and a luminosity of 9$\times10^5 M_{\odot}$,
$S=1.2$, which is almost exactly what one would expect from the Pritzl
et al.\ (2003) relations --- log $S$ both as a function of absolute
magnitude and as a function of [Fe/H].

\section{Summary and conclusions}

A search for variable stars has been conducted in the Phoenix dwarf
galaxy. Nineteen Cepheids (either AC or s-pCC), six candidate
long-period variables, one candidate eclipsing binary and a large
number of candidate RR Lyrae stars have been identified. Periods and
light curves have been obtained for all the Cepheid variables. Their
distribution in the PL diagram reveals that both AC and s-pCC are
present in our sample. This is the first time that both types of
variable star, which belong to metal-poor old/intermediate-age and
metal-poor young populations respectively, have been identified in the
same system. We show, however, that they also likely coexist in Leo~A
and Sextans~A, even thought the fact originally went unnoticed. We
note that, thanks to the very specific conditions of age and
metallicity required for the occurrence of these variables, they can
provide important hints on the age--metallicity relation of the host
galaxy. For example, in the case of Phoenix they imply, according to
current stellar-evolution models, a very low metallicity ($Z=0.0001$)
for intermediate-age ($\simeq$ 1--2 Gyr) stars, while a slightly
larger metallicity $Z=0.0004-0.001$ is possible for the young ($<
1\,$Gyr) population in the galaxy.

The RR Lyrae candidates, together with the well developed horizontal
branch, trace the existence of an important old population in Phoenix.
The different spatial distributions of AC, s-pCC, and RR Lyrae
variables is consistent with the stellar population gradients found in
Phoenix (MGA99), in the sense that the younger populations are
progressively more concentrated toward the central part of the
galaxy. The gradient in the mean age of the youngest populations in
the center of Phoenix, which seems to indicate a propagation of the
recent star formation as suggested by MGA99, is also reflected in the
spatial distribution of the s-pCC.

\acknowledgments

We want to thank M. Wischnjewsky\footnote{Deceased 2002 September} \&
O. Pevunova for their help in the data reduction, and G. Bono for
useful discussions. C.G. wants to thank J.M. G\'omez-Forrellad,
A.K. Vivas and P. Rodr\'\i guez-Gil for obtaining periods of some
variables stars at early stages of this work, while she was developing
her own code, and G. Bono for useful discussions. C.G. acknowledges
partial support from the Spanish Ministry of Science and Technology
(Plan Nacional de Investigaci\'on Cient\' \i fica, Desarrollo e
Investigaci\'on Tecnol\'ogica, AYA2002-01939), and from the European
Structural Funds.


\clearpage

\clearpage

\clearpage

\begin{figure}
\epsscale{1.0}
\figurenum{3}
\plottwo{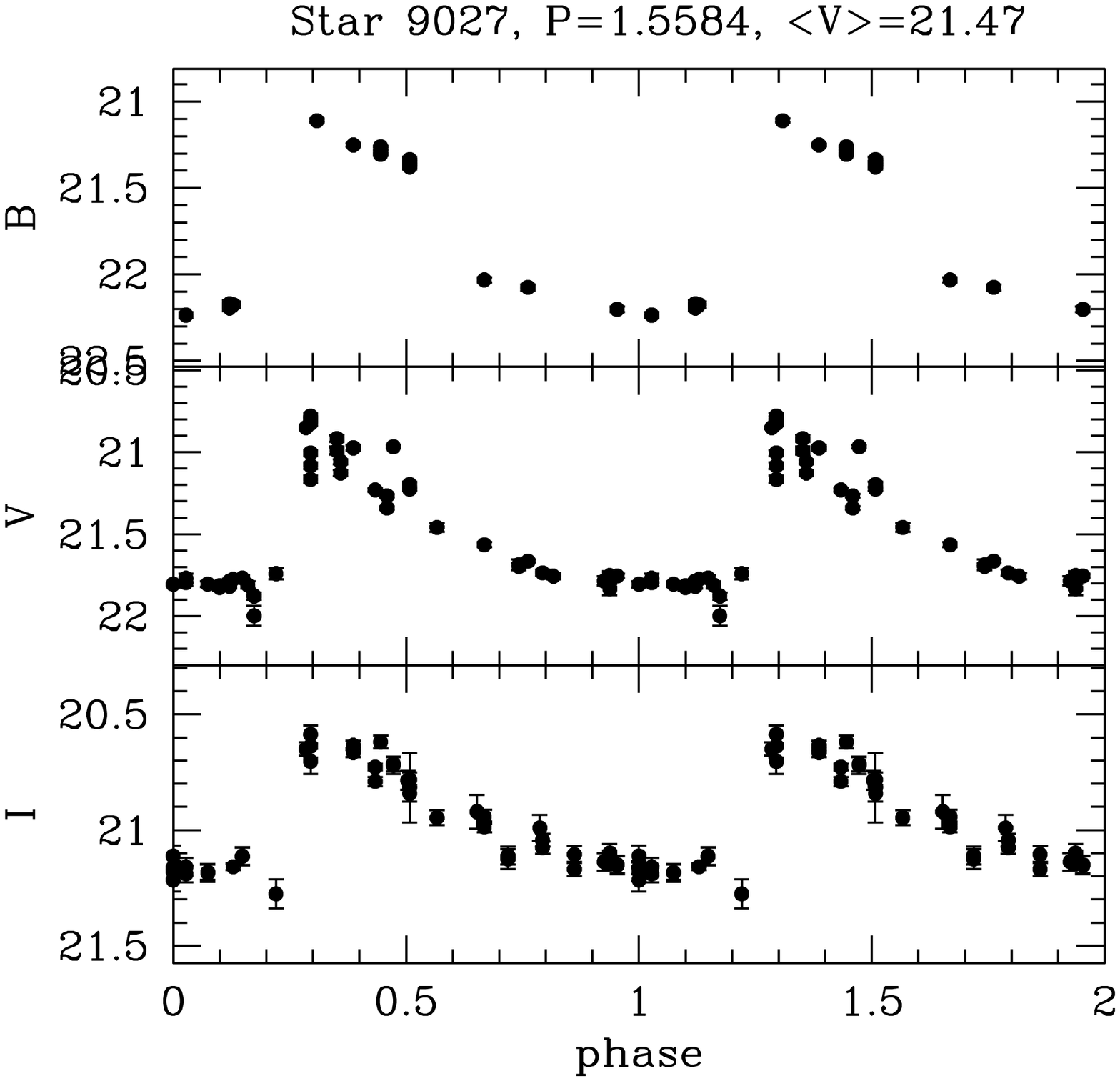}{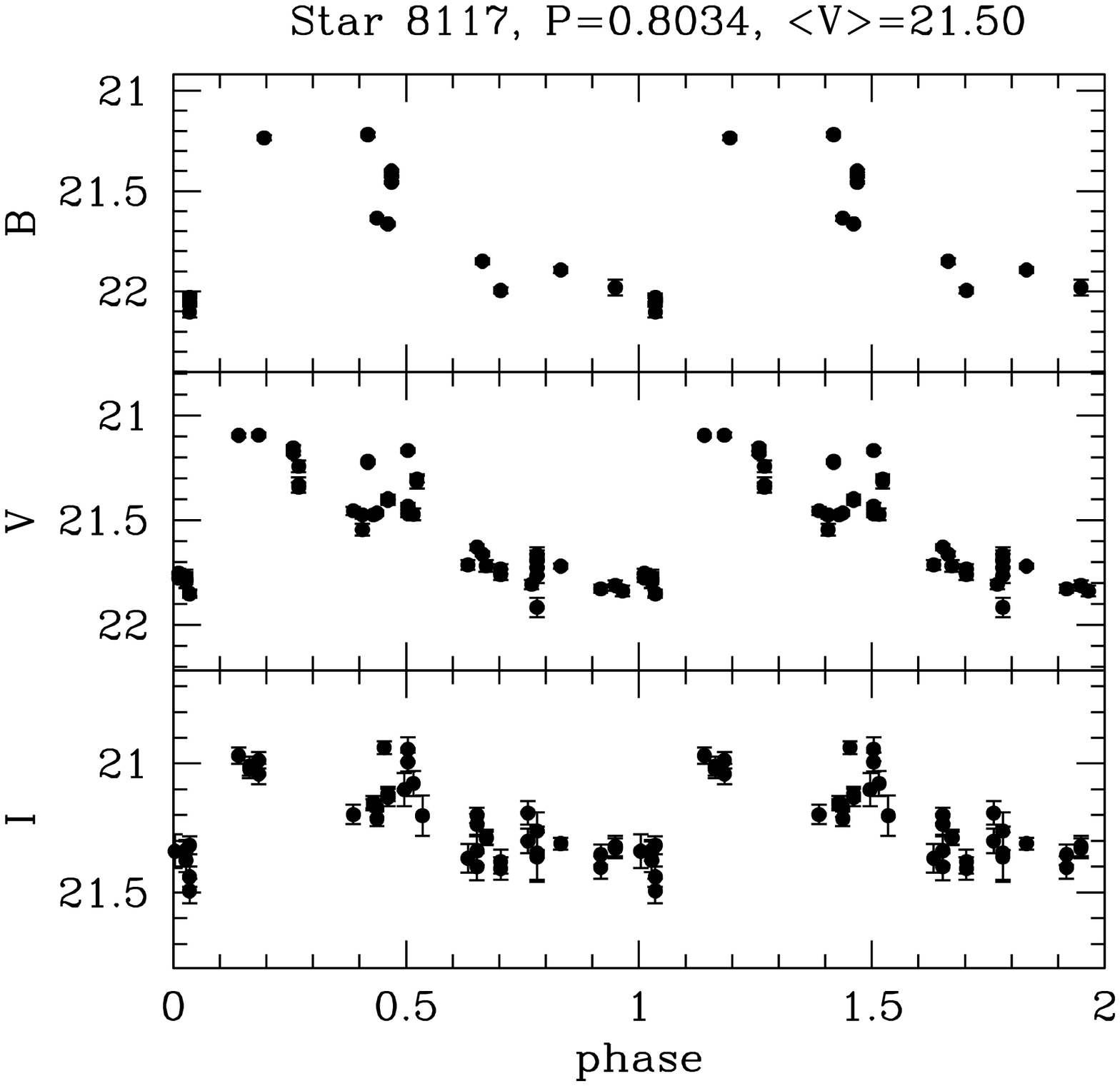}
\caption{b)}
\end{figure} 
\clearpage

\begin{figure}
\figurenum{3}
\plottwo{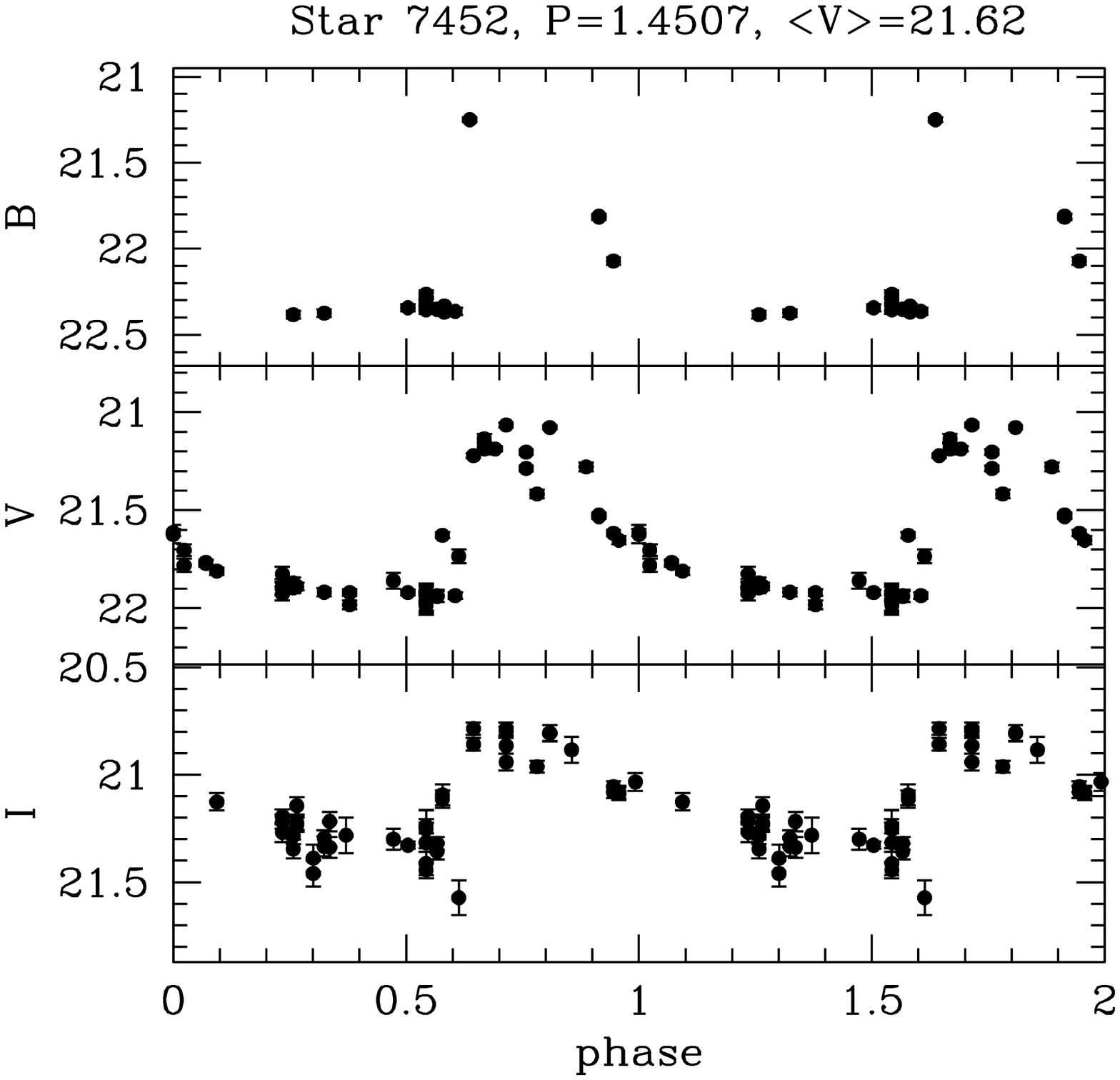}{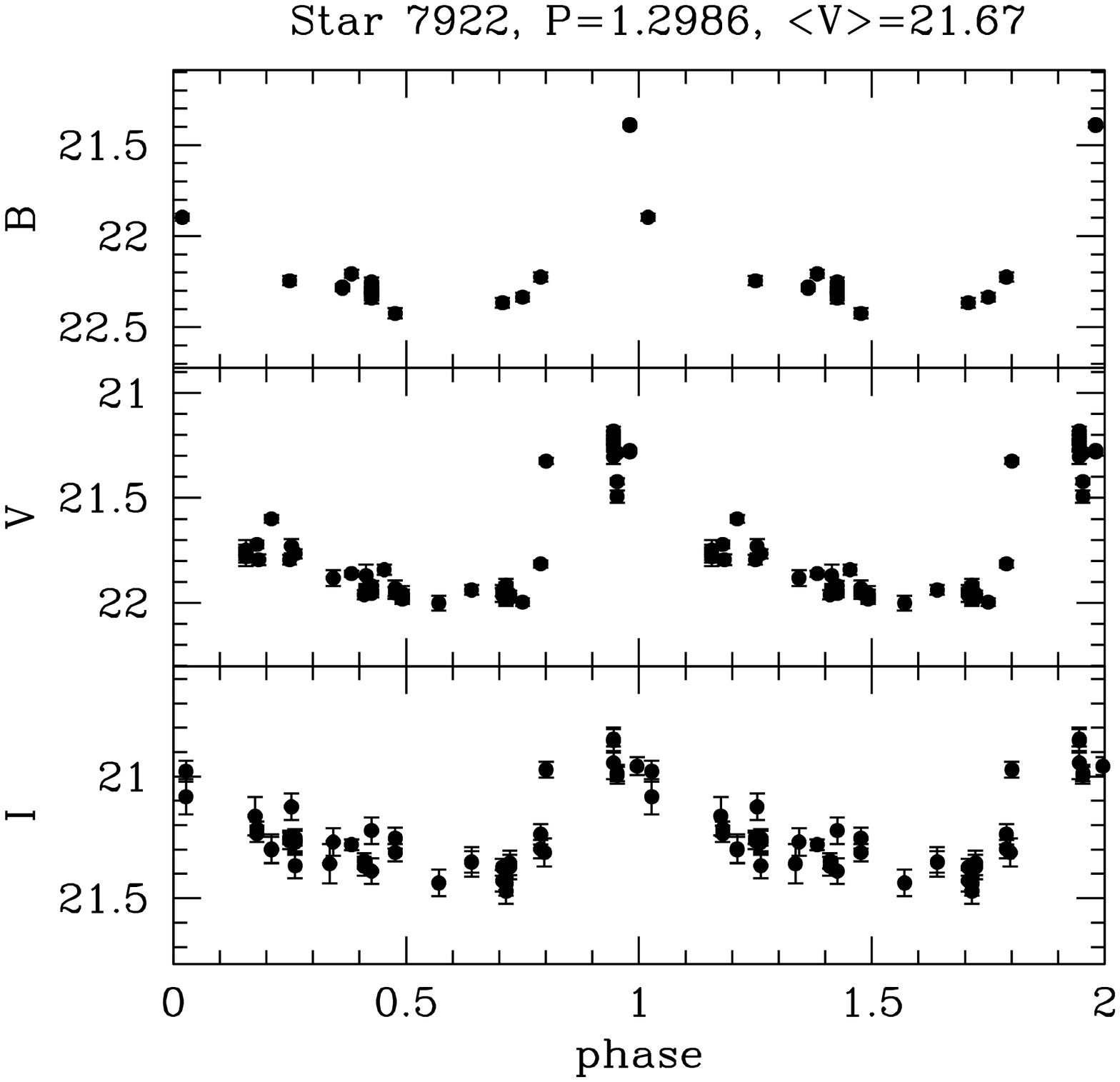}
\caption{c)}
\end{figure} 

\begin{figure}
\figurenum{3}
\plottwo{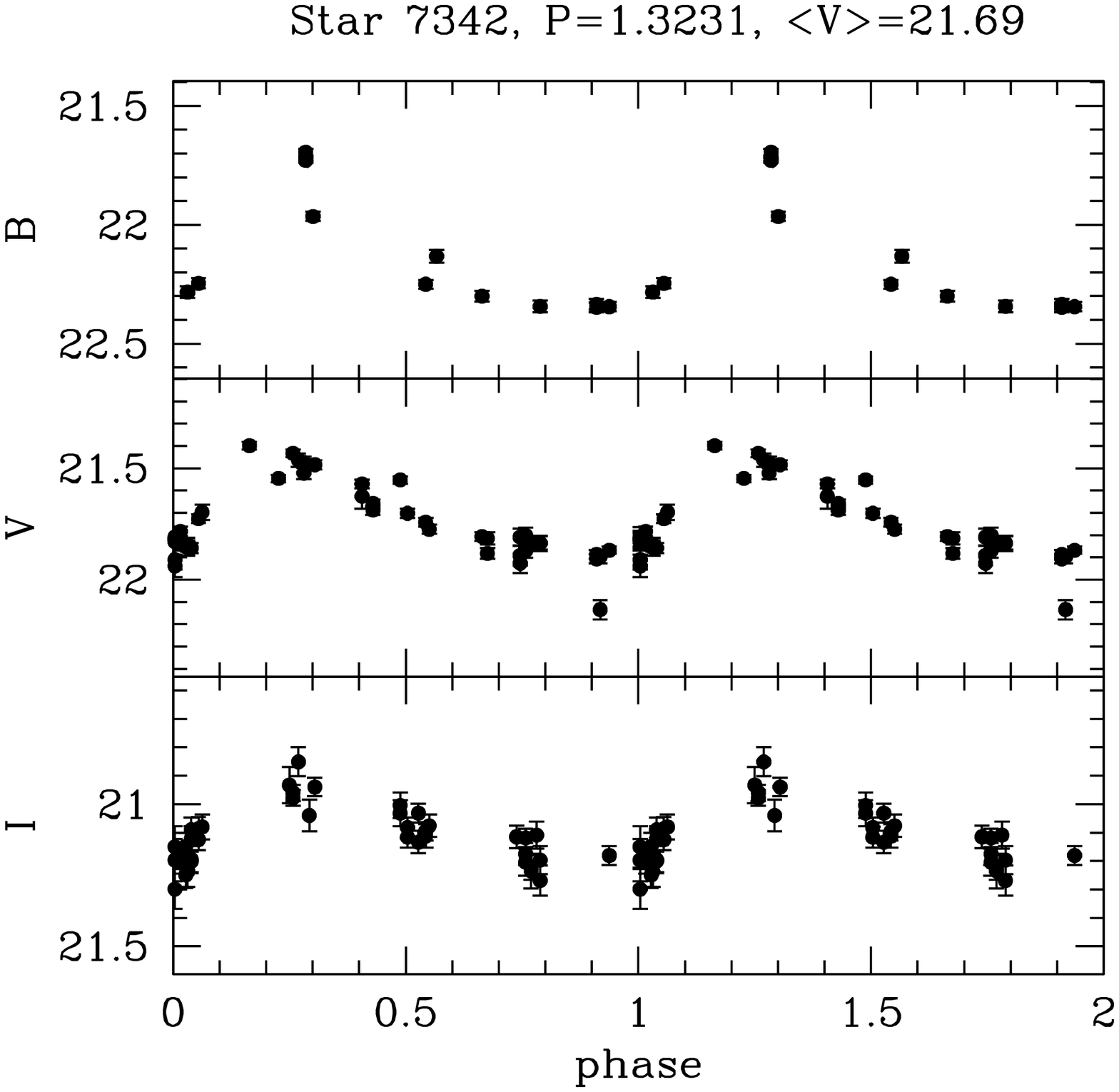}{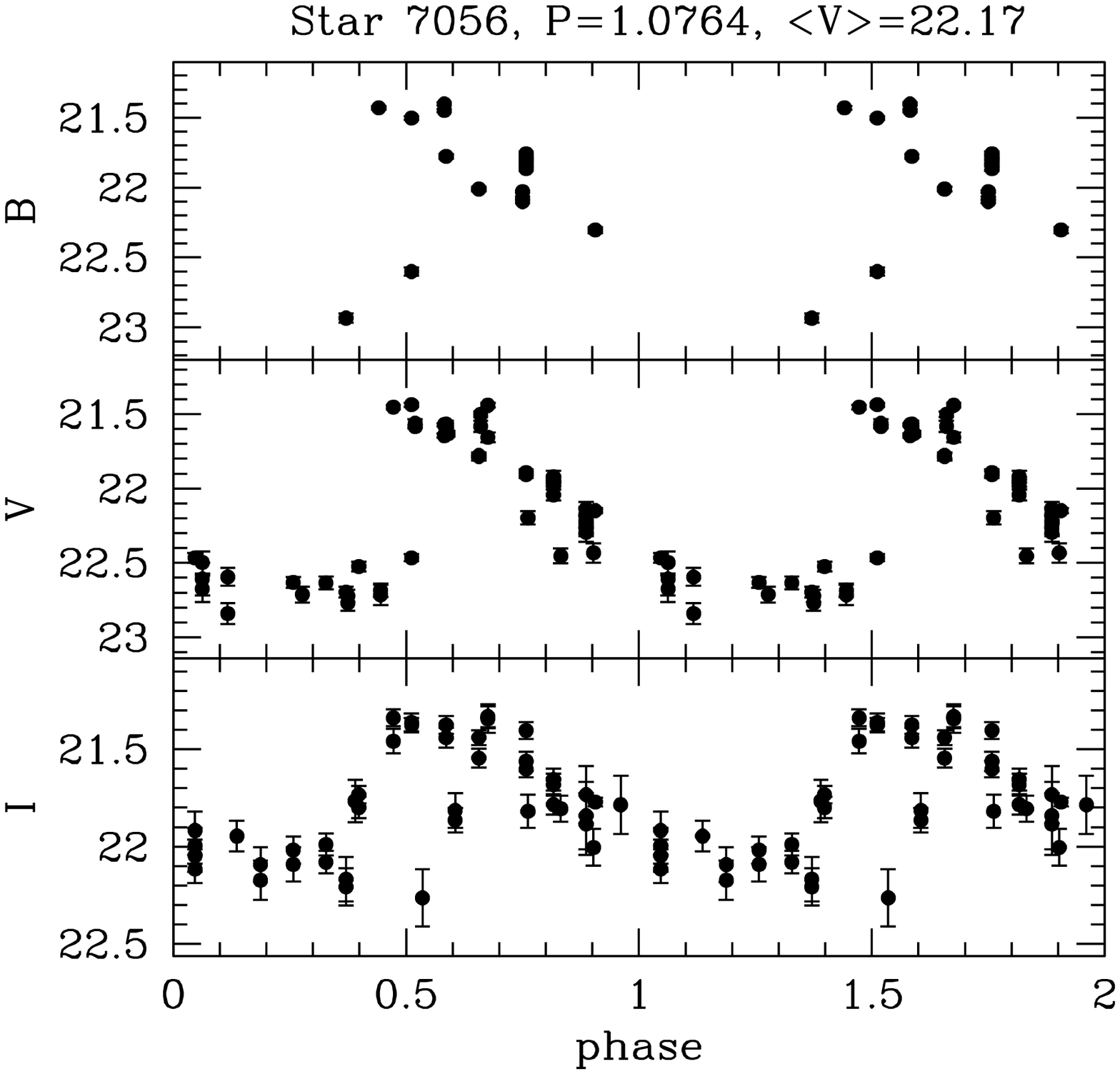}
\caption{d)}
\end{figure} 

\begin{figure}
\figurenum{3}
\plottwo{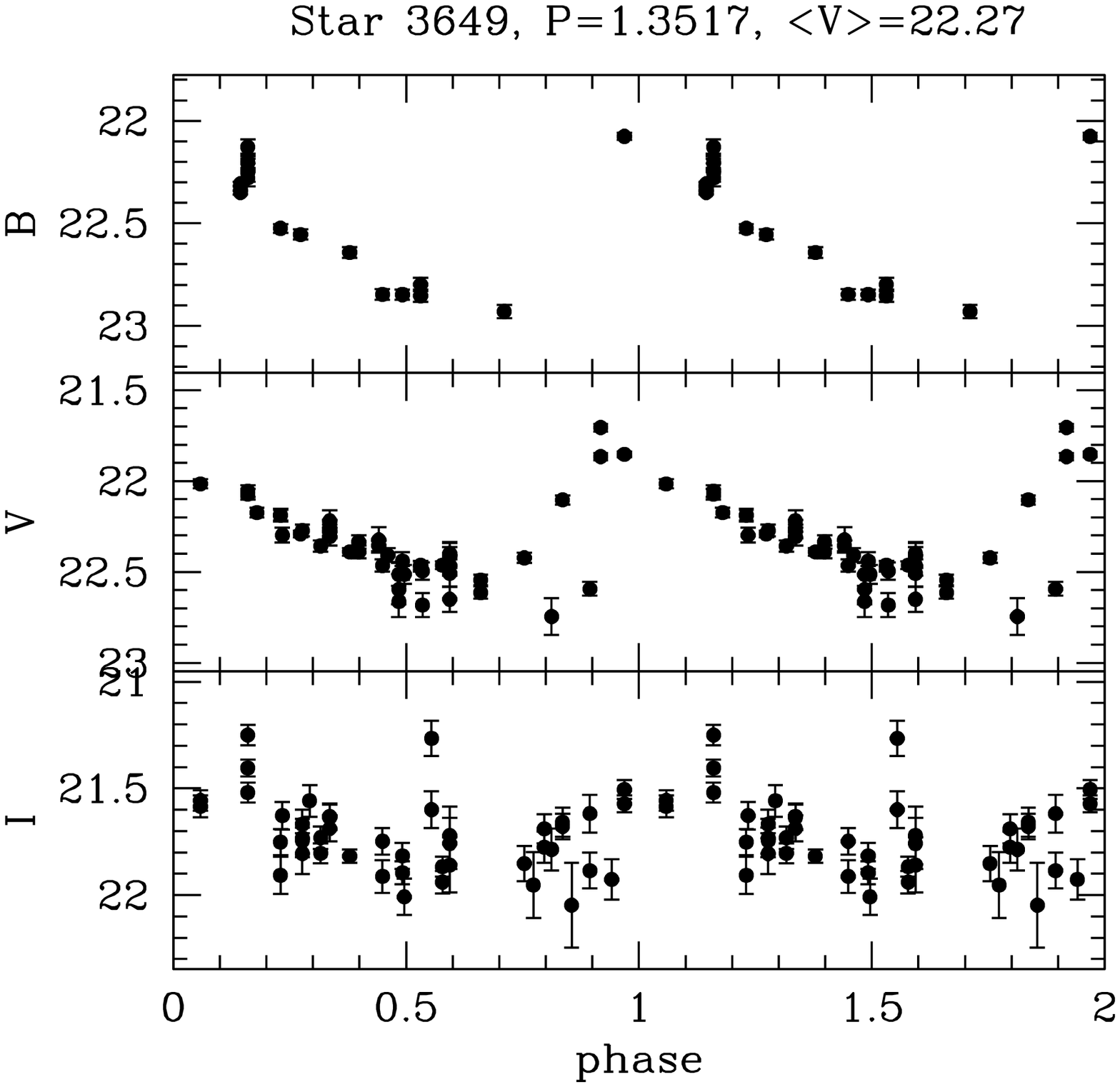}{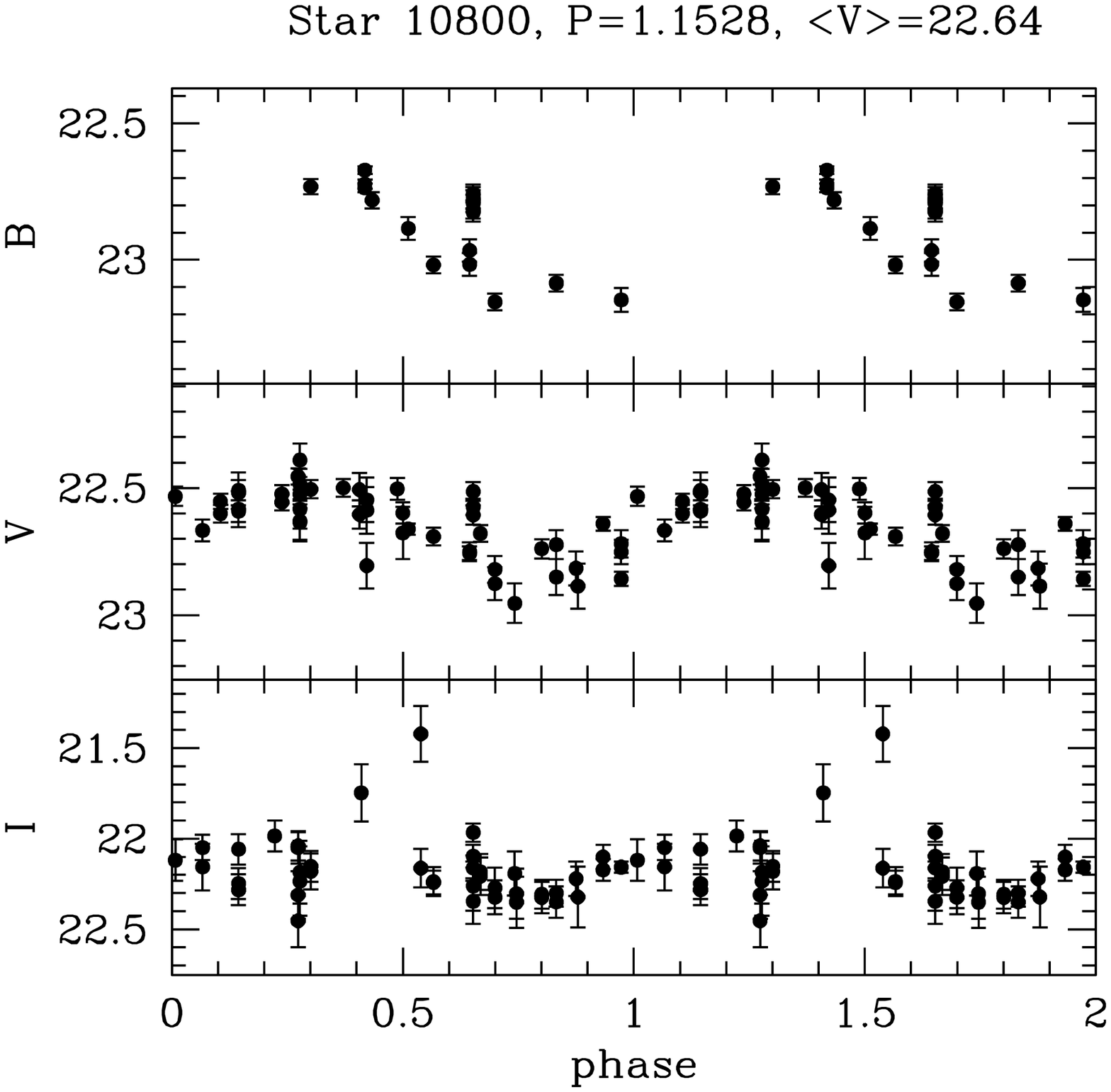}
\caption{e)}
\end{figure} 

\begin{figure}
\figurenum{3}
\plottwo{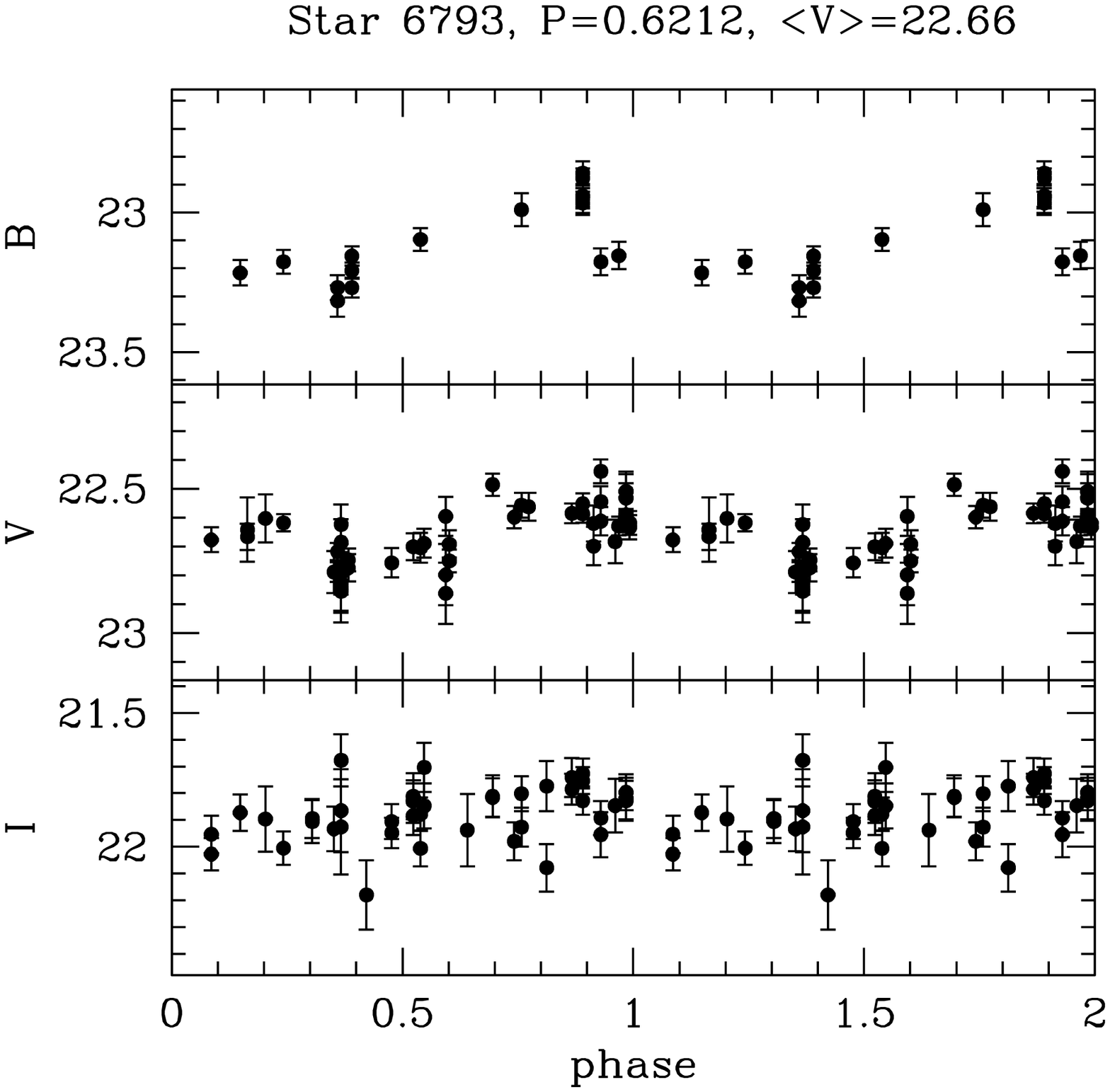}{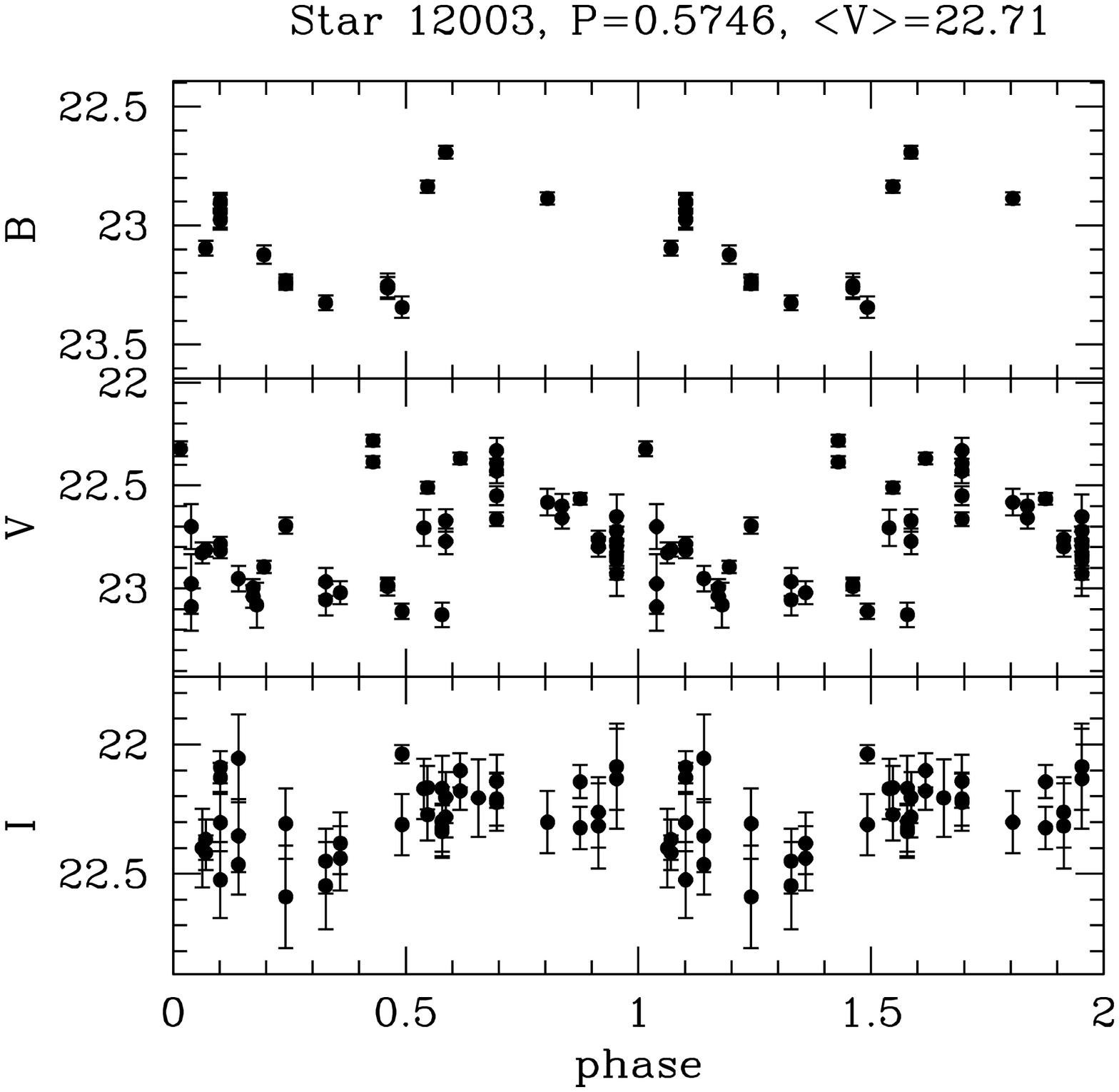}
\caption{f)}
\end{figure} 
\clearpage

\begin{figure}
\figurenum{3}
\plottwo{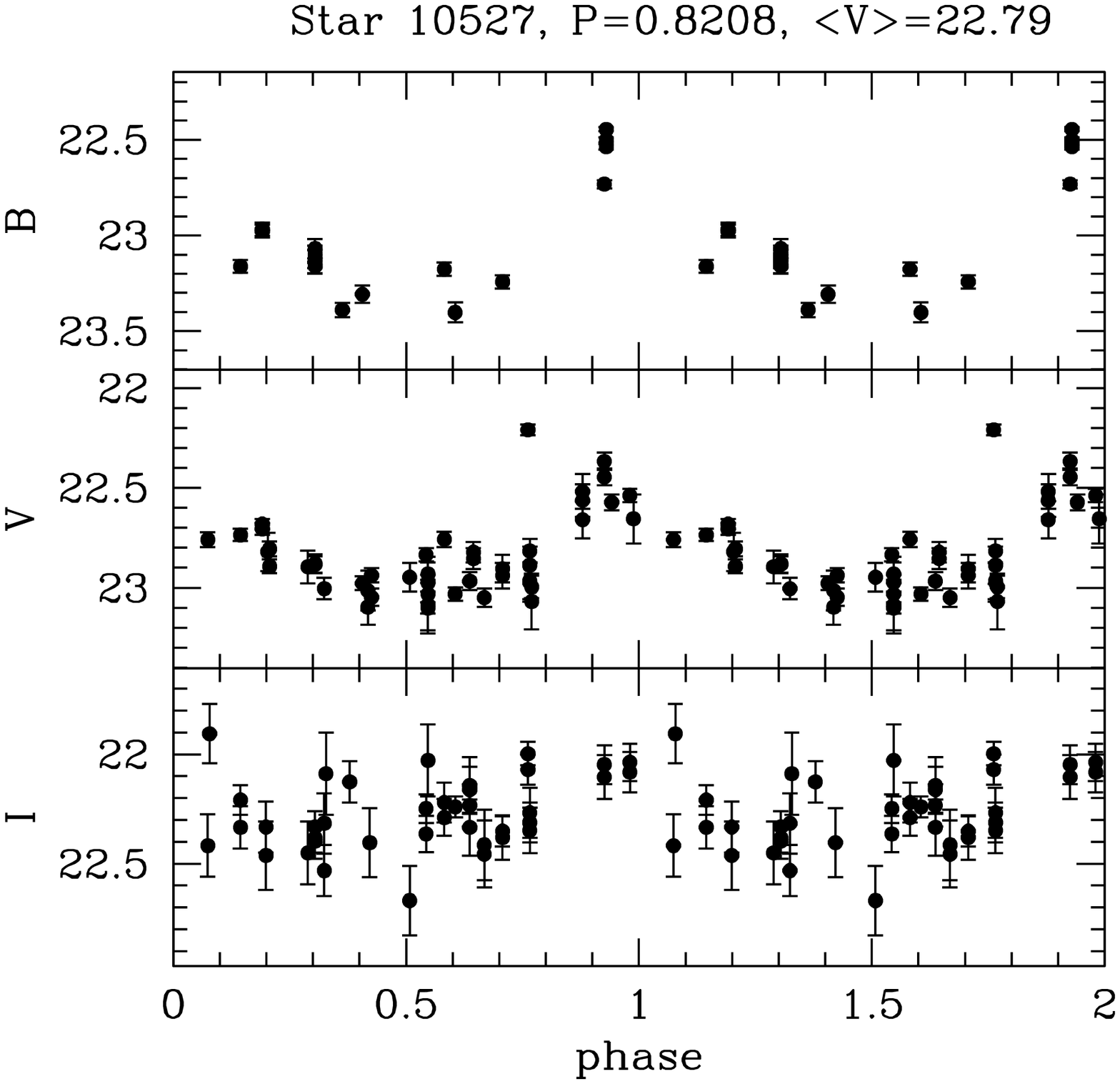}{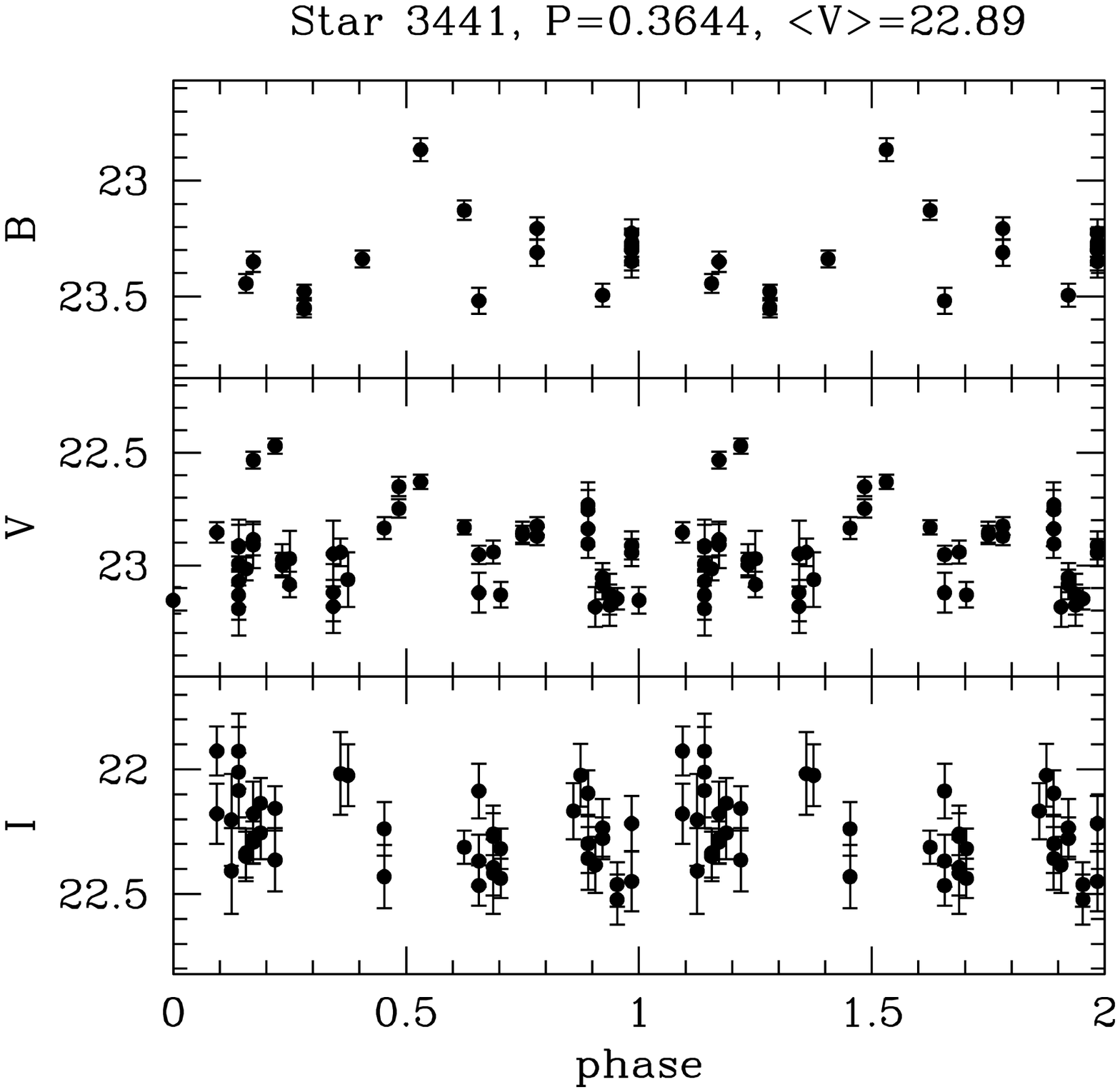}
\caption{g)}
\end{figure} 

\begin{figure}
\figurenum{3}
\plottwo{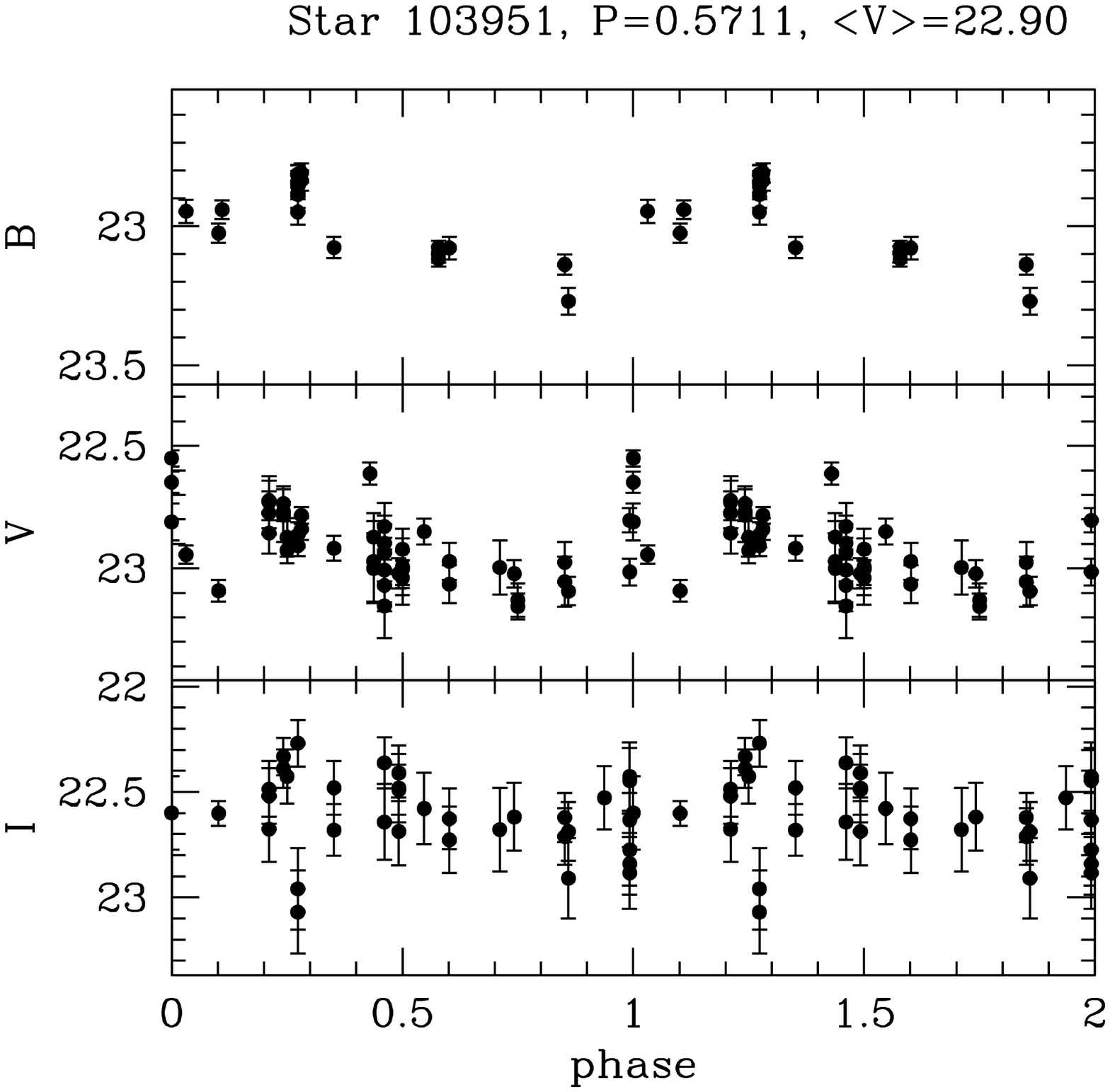}{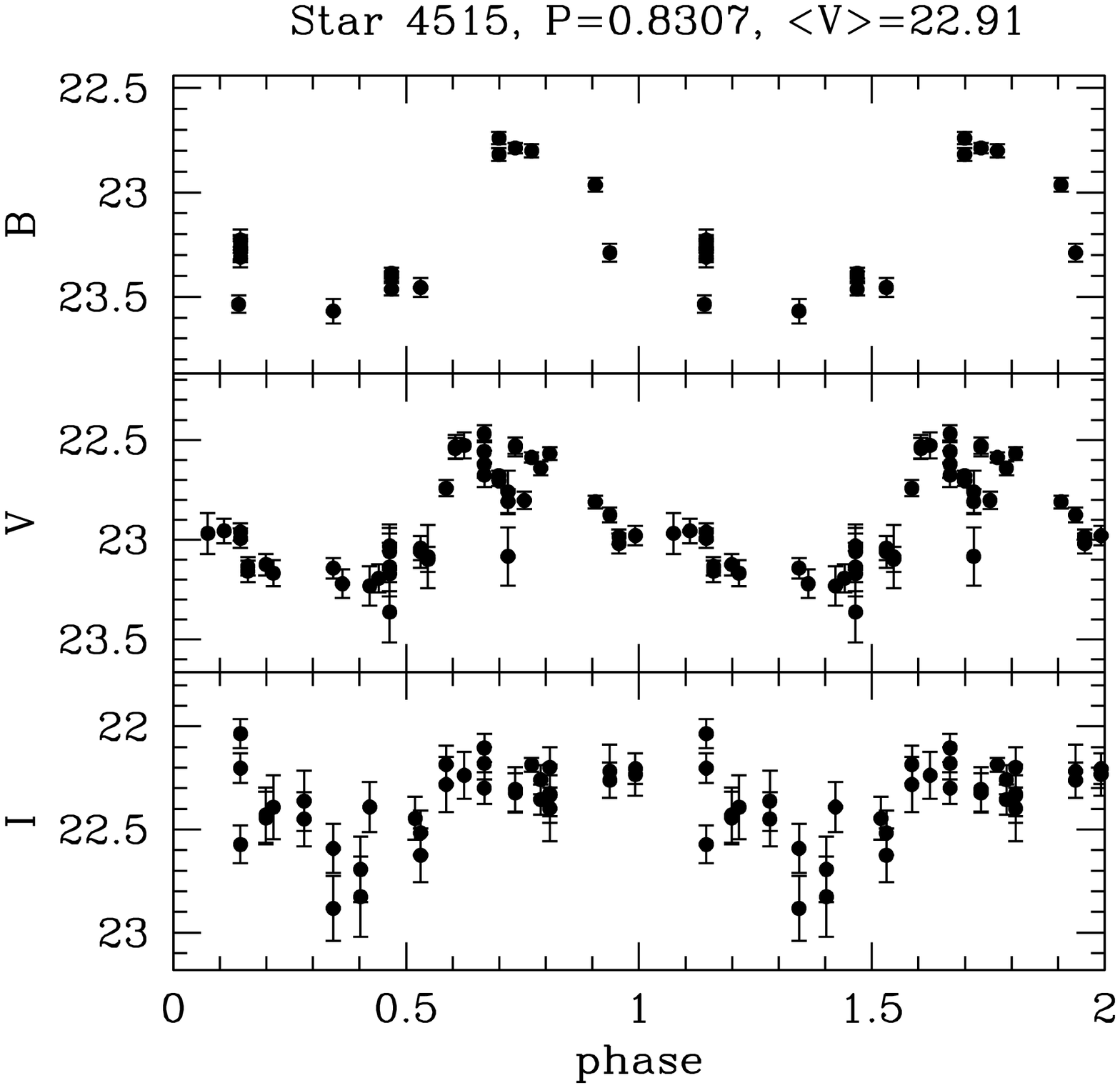}
\caption{h)}
\end{figure}

\clearpage

\begin{figure}
\figurenum{3}
\plottwo{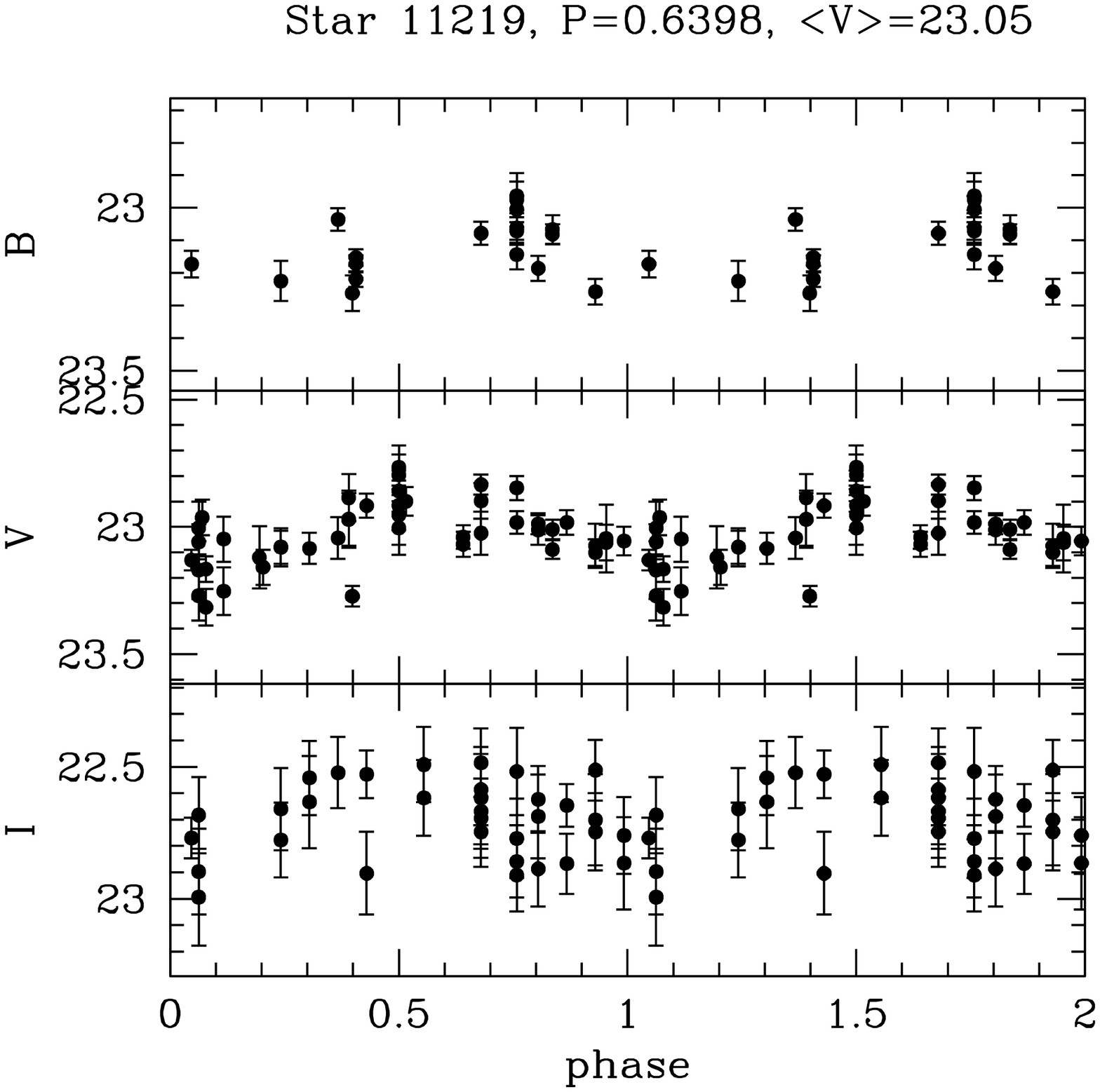}{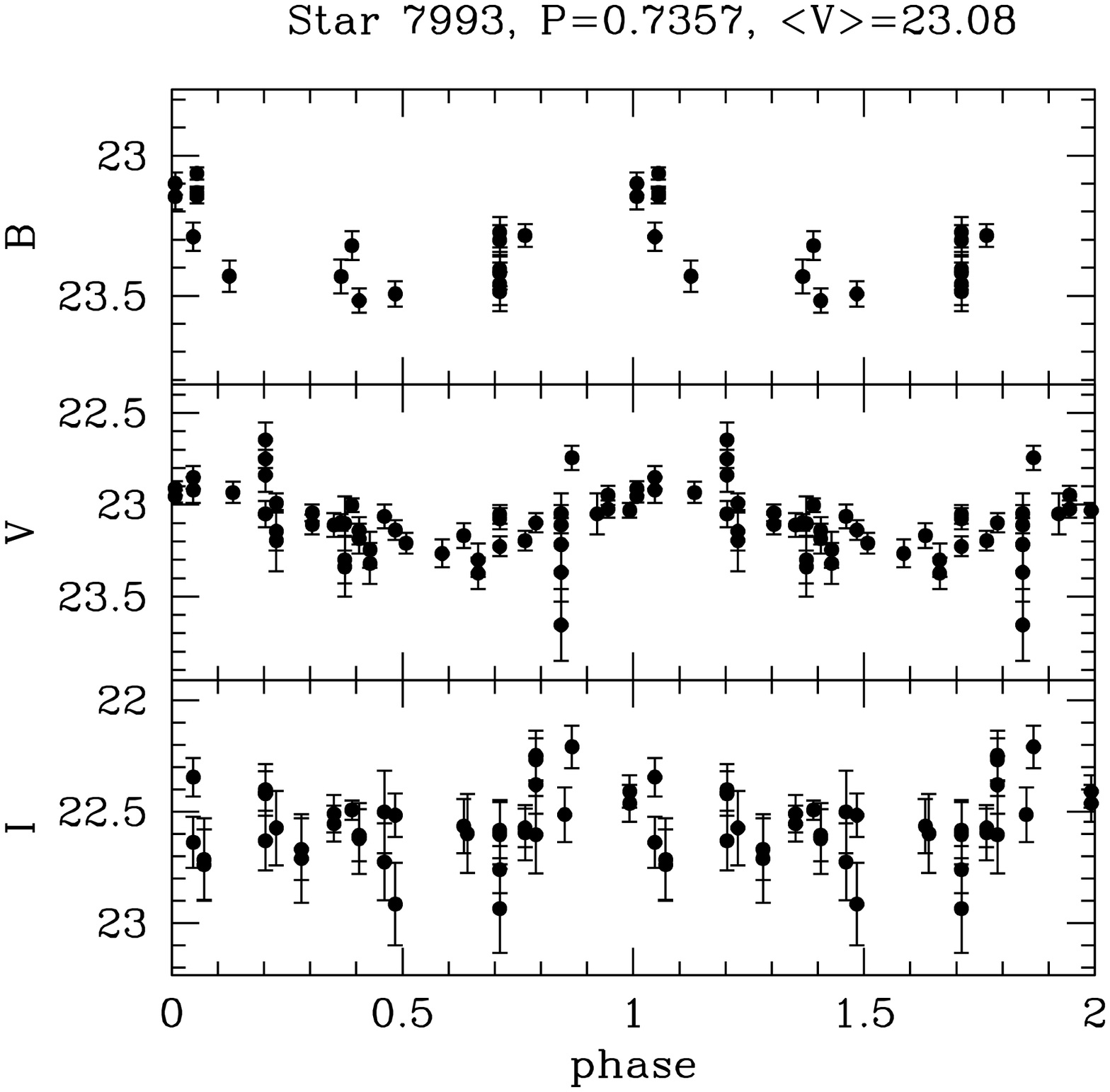}
\caption{i)}
\end{figure}


\begin{figure}
\figurenum{3}
\plotone{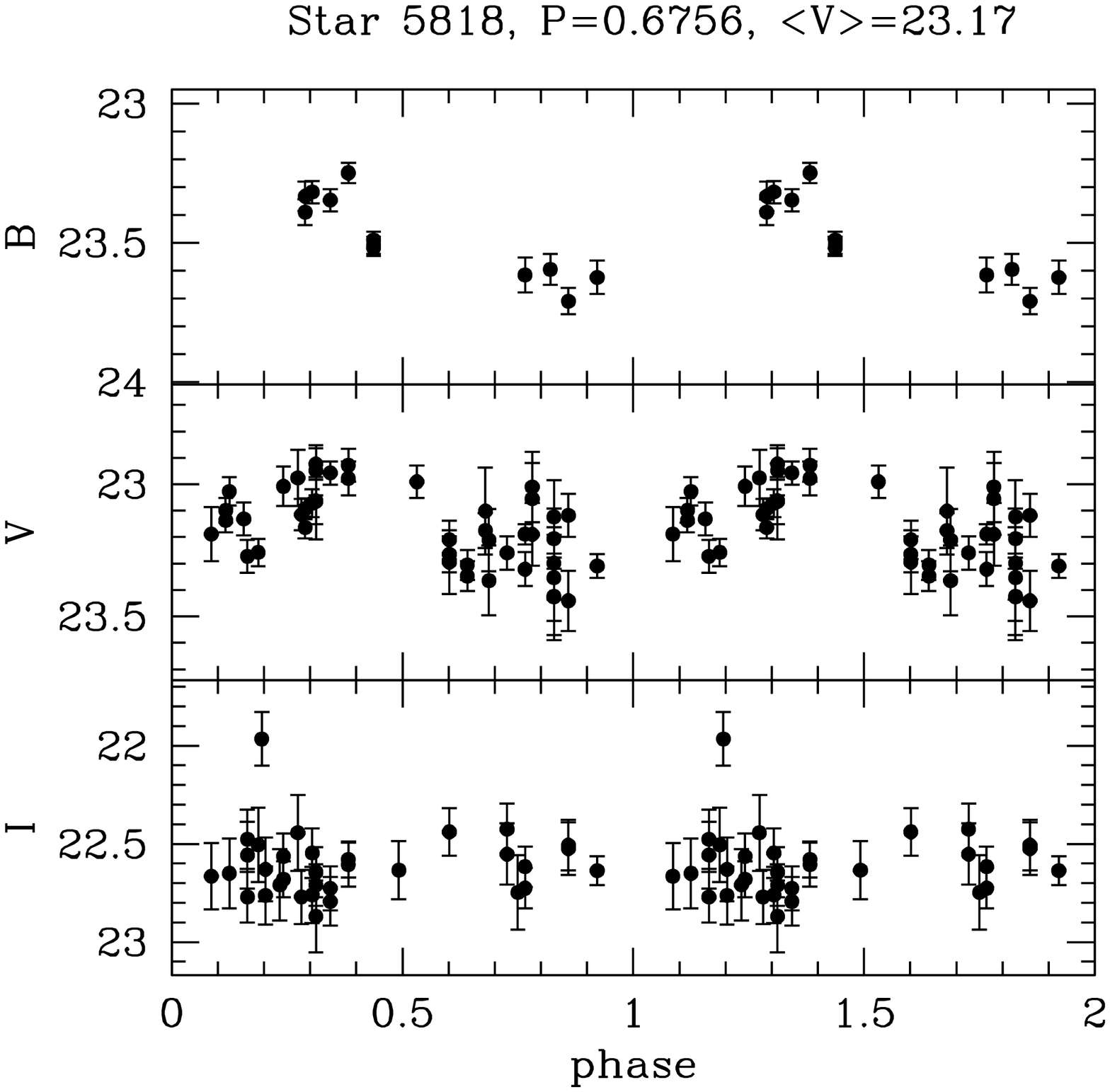}
\caption{j)}
\end{figure}

\clearpage

\begin{figure}
\plotone{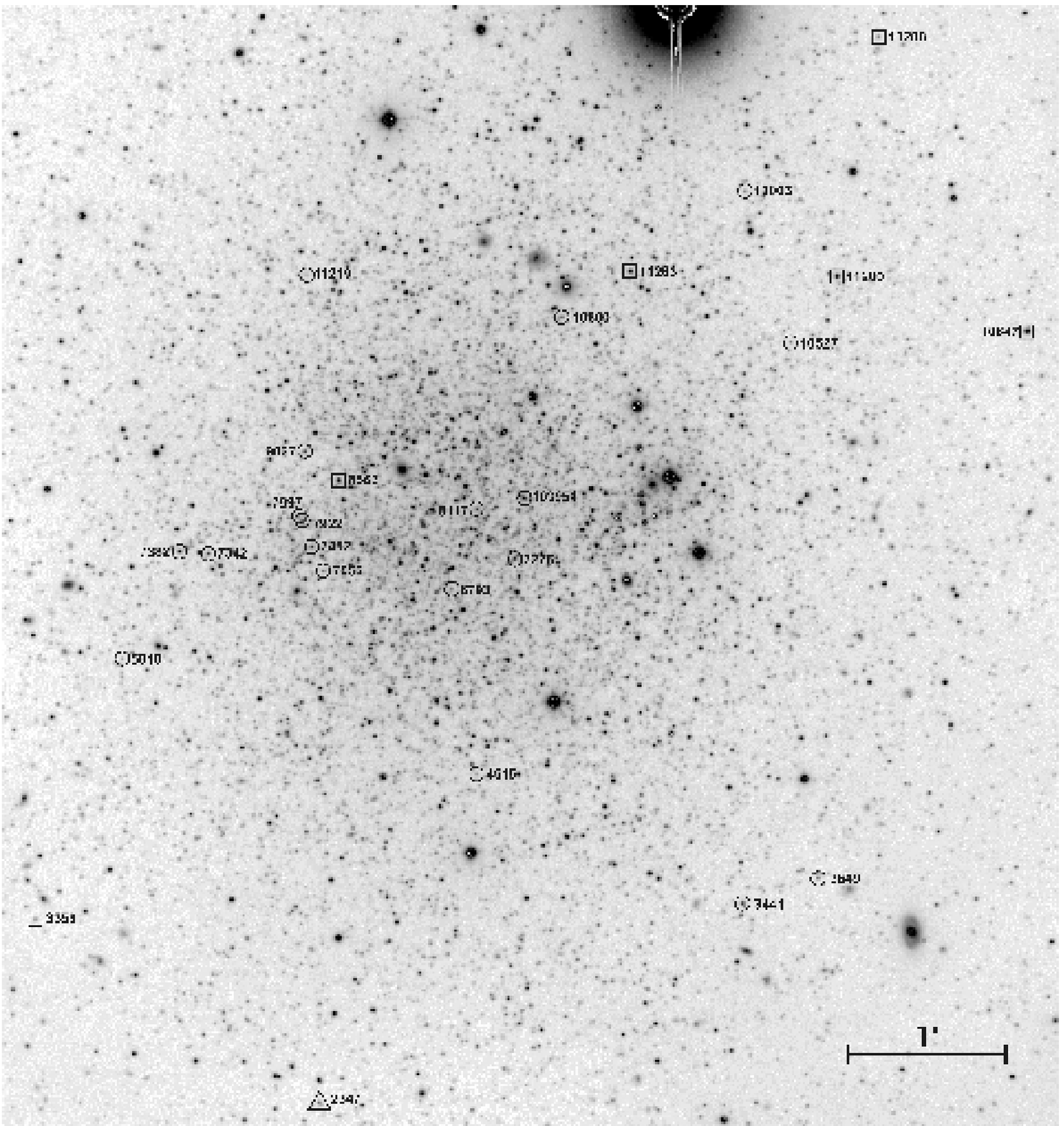}
\caption{Finder chart for the Phoenix variable stars. Confirmed Cepheid variables are marked with circles, candidate long-period variables with squares, and the candidate eclipsing binary with a triangle. North is down and east is to the left.}
\label{mapas}
\end{figure} 
\clearpage

\begin{figure}
\plotone{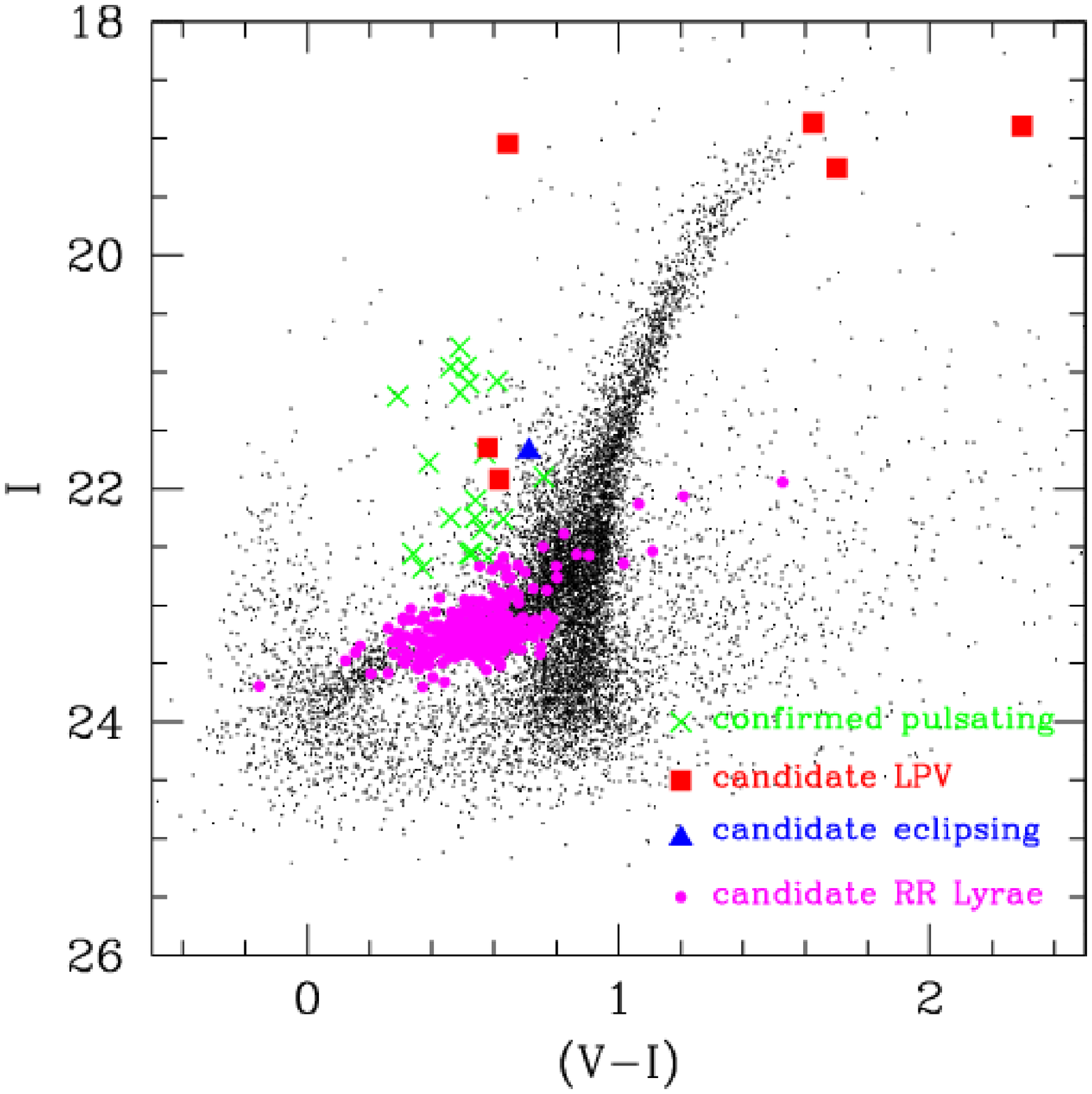}
\caption{Confirmed pulsating (crosses) and long-period (squares) variables, and 
the candidate eclipsing binary star (triangle). The 396 candidate RR Lyrae 
stars are shown as small circles.} 
\label{dcmvilim}
\end{figure}
\clearpage

\begin{figure}
\plottwo{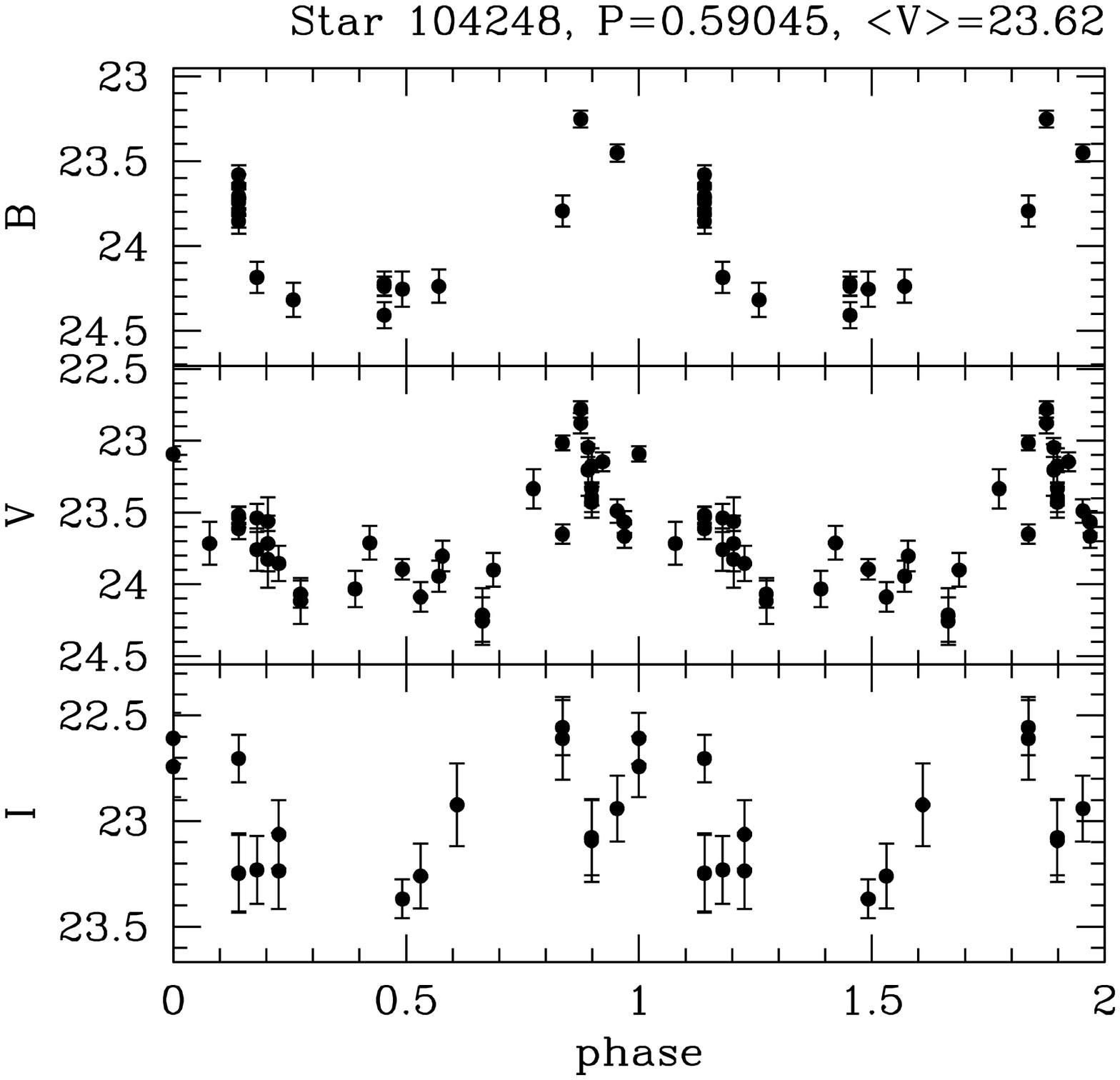}{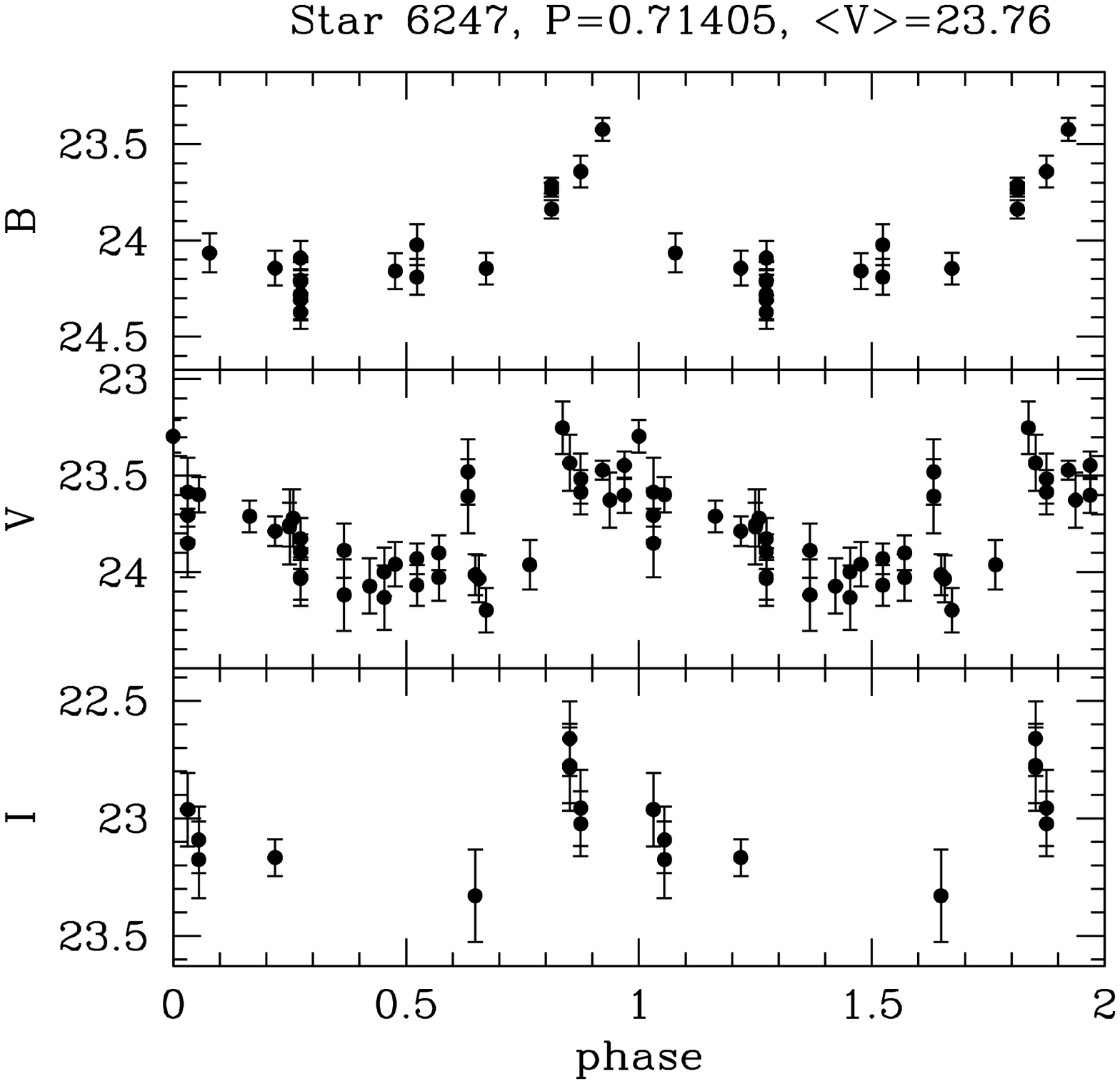}
\caption{a) $B$, $V$ and $I$ light curves for a sample of Phoenix confirmed RR Lyrae. Not all the RR Lyrae candidates have been analyzed for periods. Data are repeated over a second cycle for clarity.\label{lc_rr}}
\end{figure} 


\begin{figure}
\figurenum{6}
\plottwo{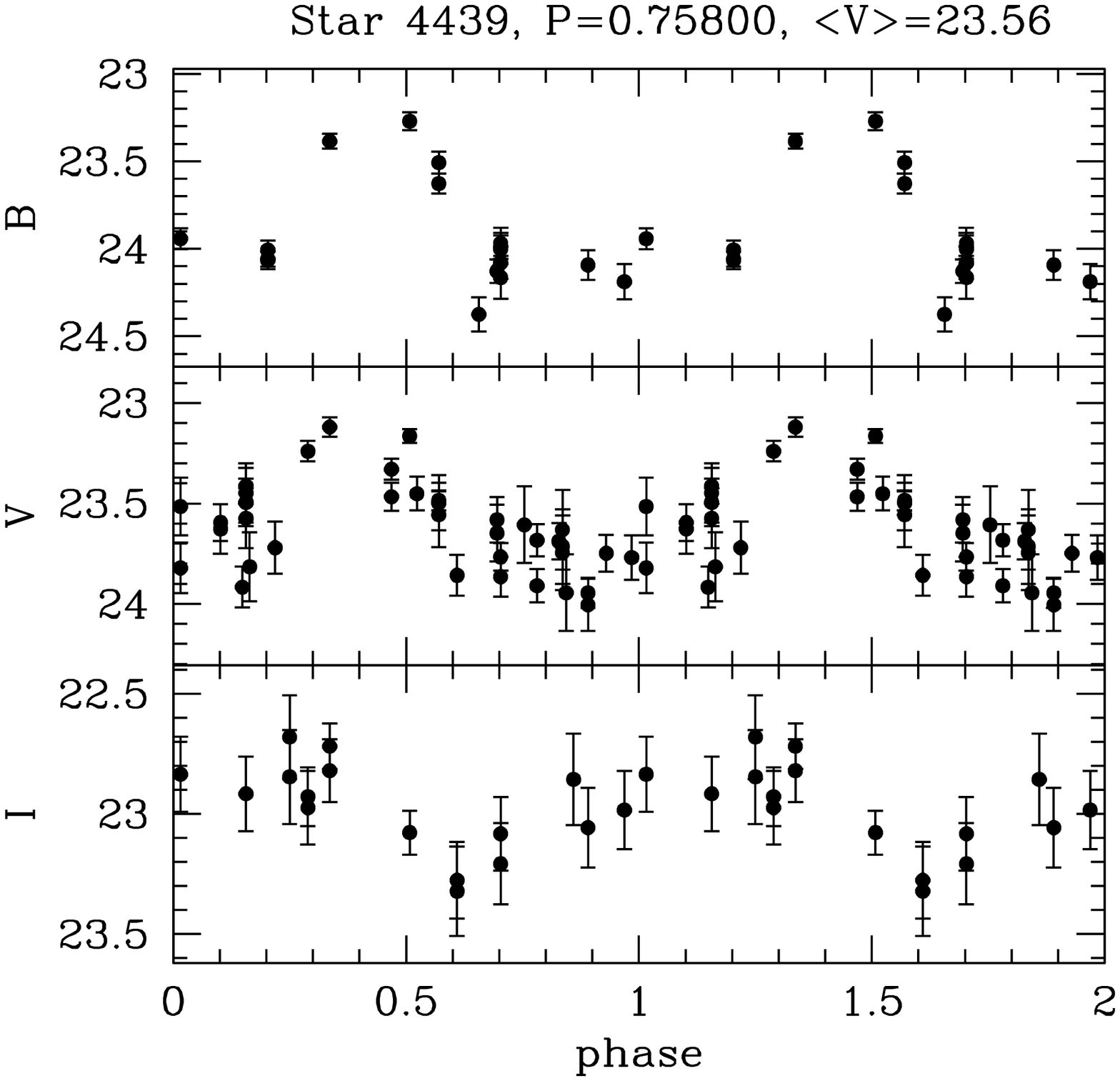}{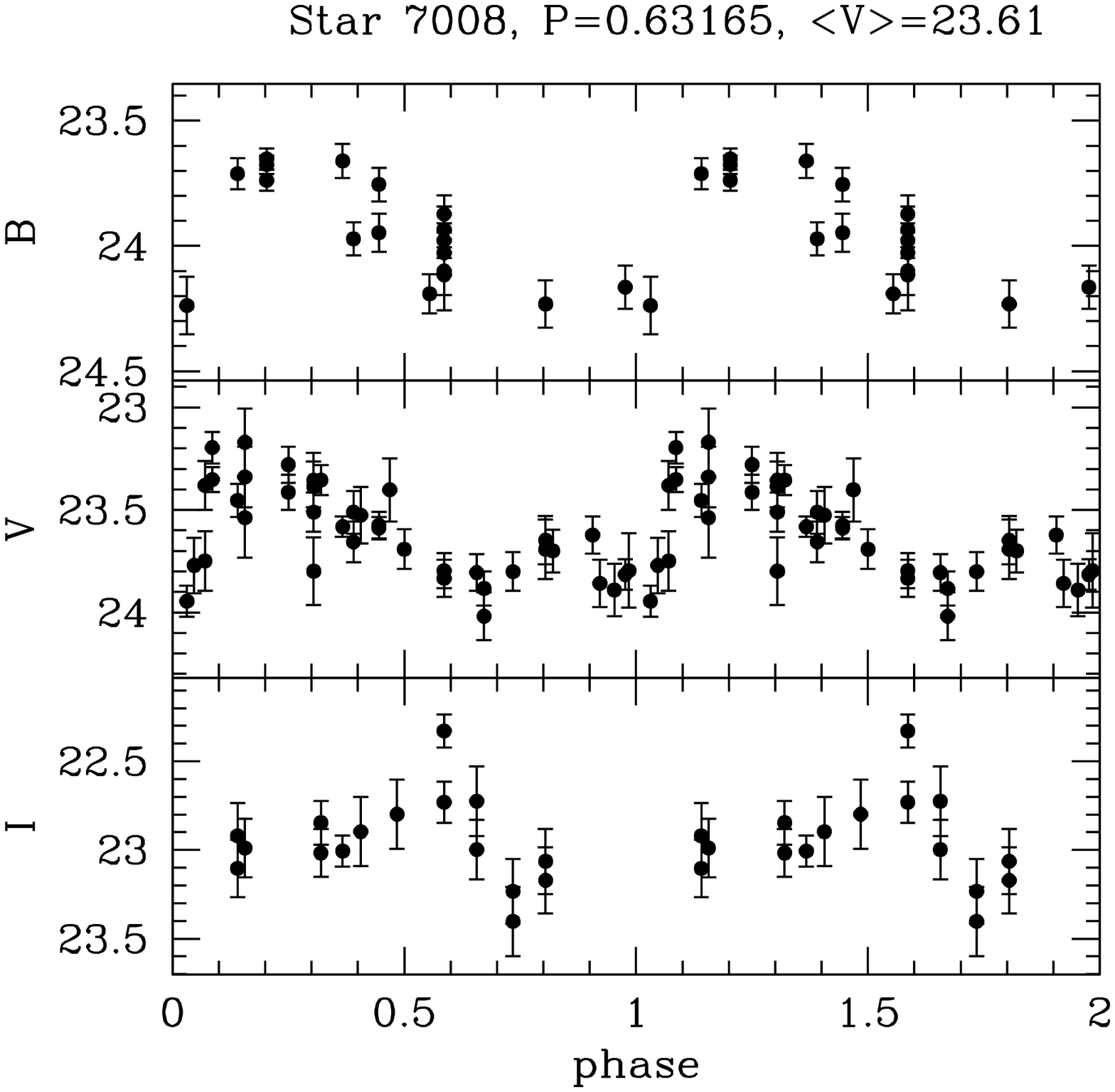}
\caption{b)}
\end{figure} 

\clearpage

\begin{figure}
\plotone{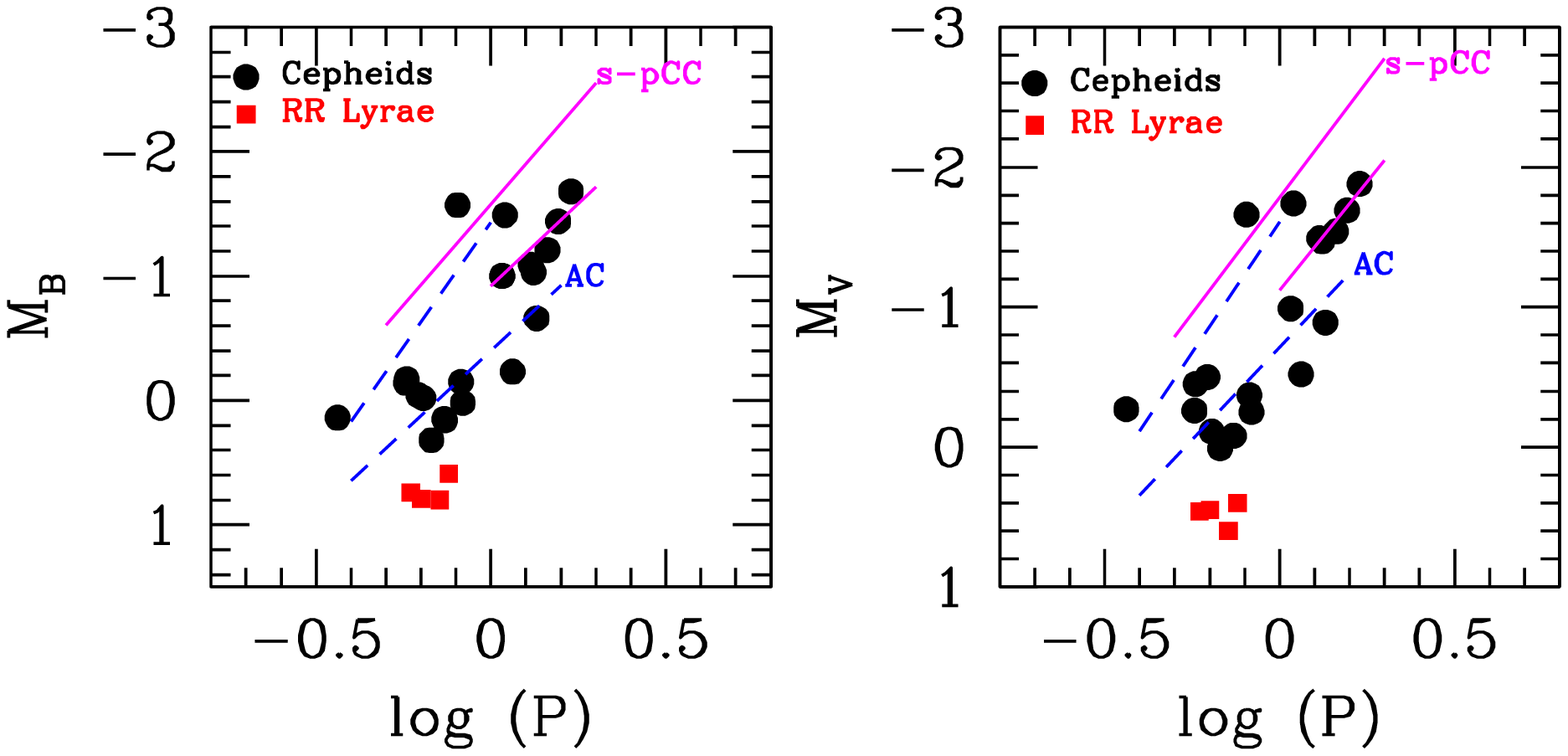}]

\caption{Period-luminosity (PL) diagrams (in the $B$ and $V$ bands) for 
the Phoenix Cepheids. The PL relations obtained for dSph AC by Pritzl
et al.\ (2002) are represented as dashed lines, and the PL relations
for (P$\le$ 2 days) OGLE SMC s-pCC as solid lines (see text for
details). The four RR Lyrae variables for which periods have been
obtained are included in the plot.}
\label{pl}
\end{figure}

\clearpage

\begin{figure}
\plotone{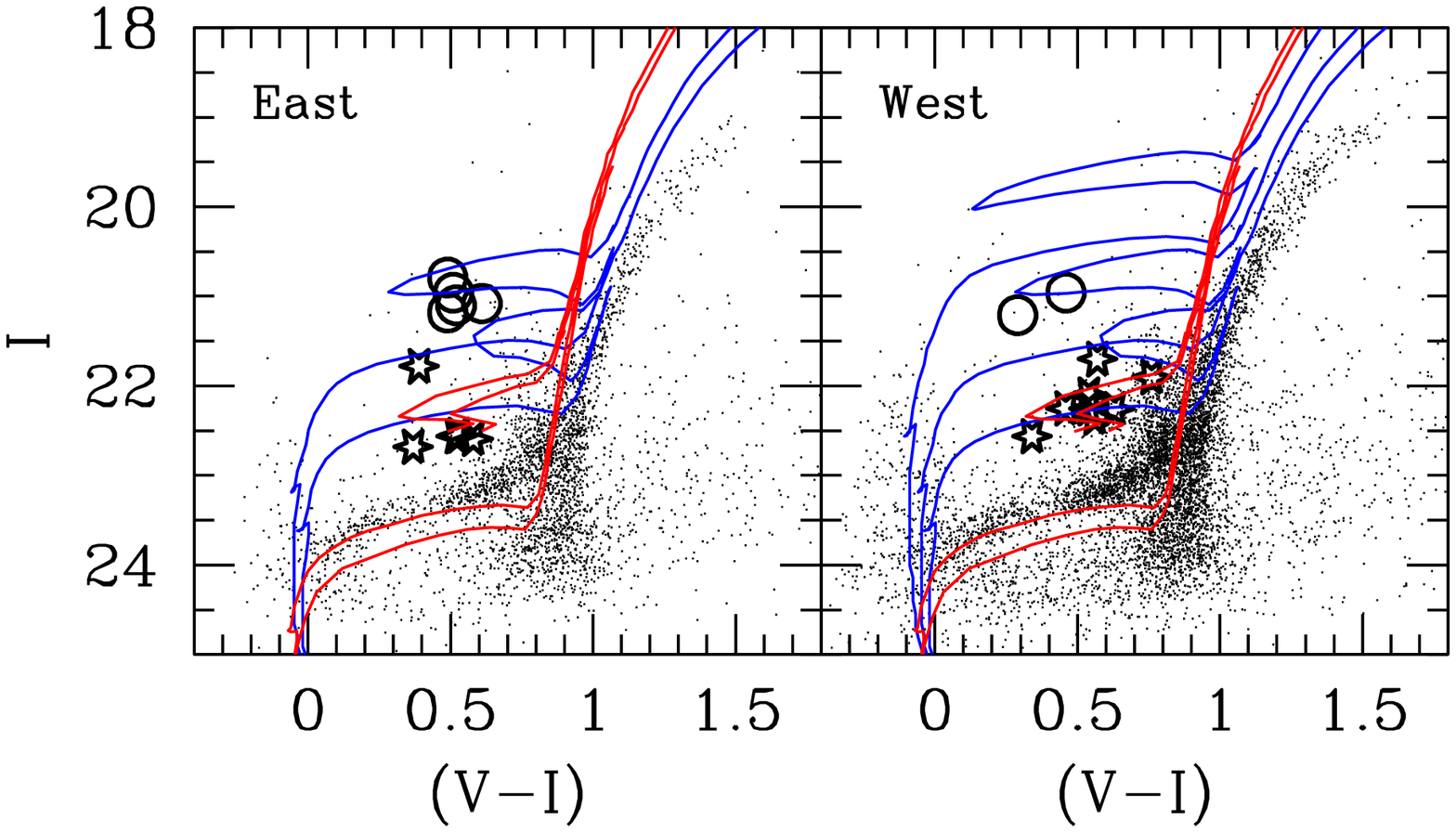}
\caption{CMDs of the eastern and western parts of Phoenix. The east/west 
division has been made in order to keep separate the stars belonging to the
young association located in the west part. It
does not divide the galaxy exactly in half, and thus the 
sequences appear more populated in the west CMD. The ratio of main sequence {\it plus}
blue-loop stars ($-0.5 \le (V-I) \le 0.1), 21 \le I \le 23.5$) to
stars near the tip of the RGB ($(V-I) \le 1.0, I \le 20$) is however
significantly larger in the west part ($1.1\pm0.2~vs~0.4\pm0.2$),
where the blue stars also extend to brighter magnitudes, indicating
younger ages. s-pCC and AC are represented as circles and stars
respectively.  Isochrones from Bertelli et al.\ (1994) for $Z=0.001$ and
ages 200, 400 and 600 Myr, and for $Z=0.0001$ and 1.3 and 1.6 Gyr have been 
superimposed. Note how the position of the s-pCC are well fit by the
younger isochrones, while the AC are best fit by the isochrones older than
1 Gyr.}
\label{dcmviiso}
\end{figure}

\newpage

\begin{figure}
\plotone{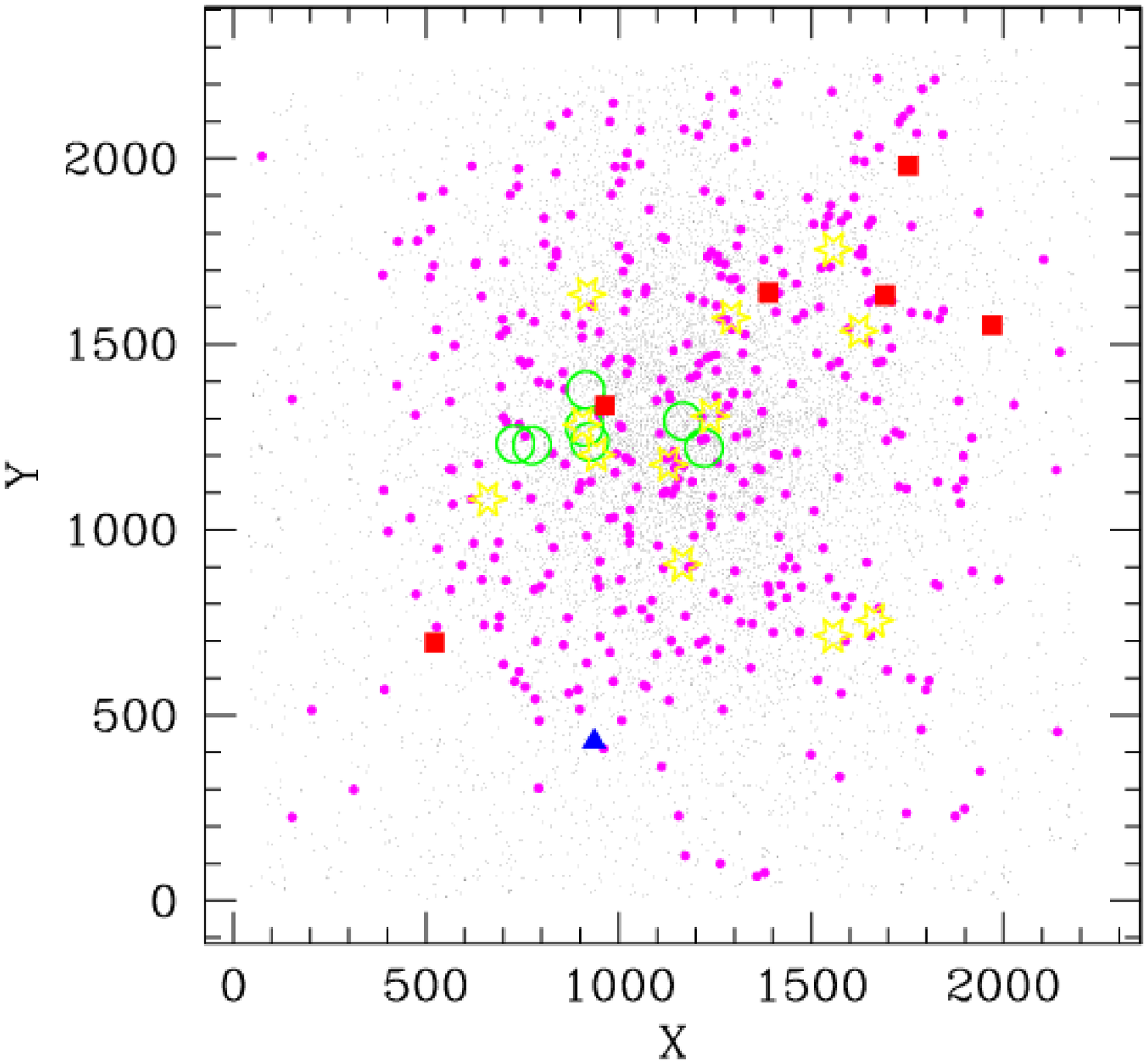}
\caption{Spatial distribution of the s-pCC (circles), AC (stars) 
long period (squares), and  the candidate eclipsing binary star 
(triangle). The 396 candidate RR Lyrae stars are shown as small 
circles.} 
\label{distrixy}
\end{figure}

\newpage

\begin{figure}
\plotone{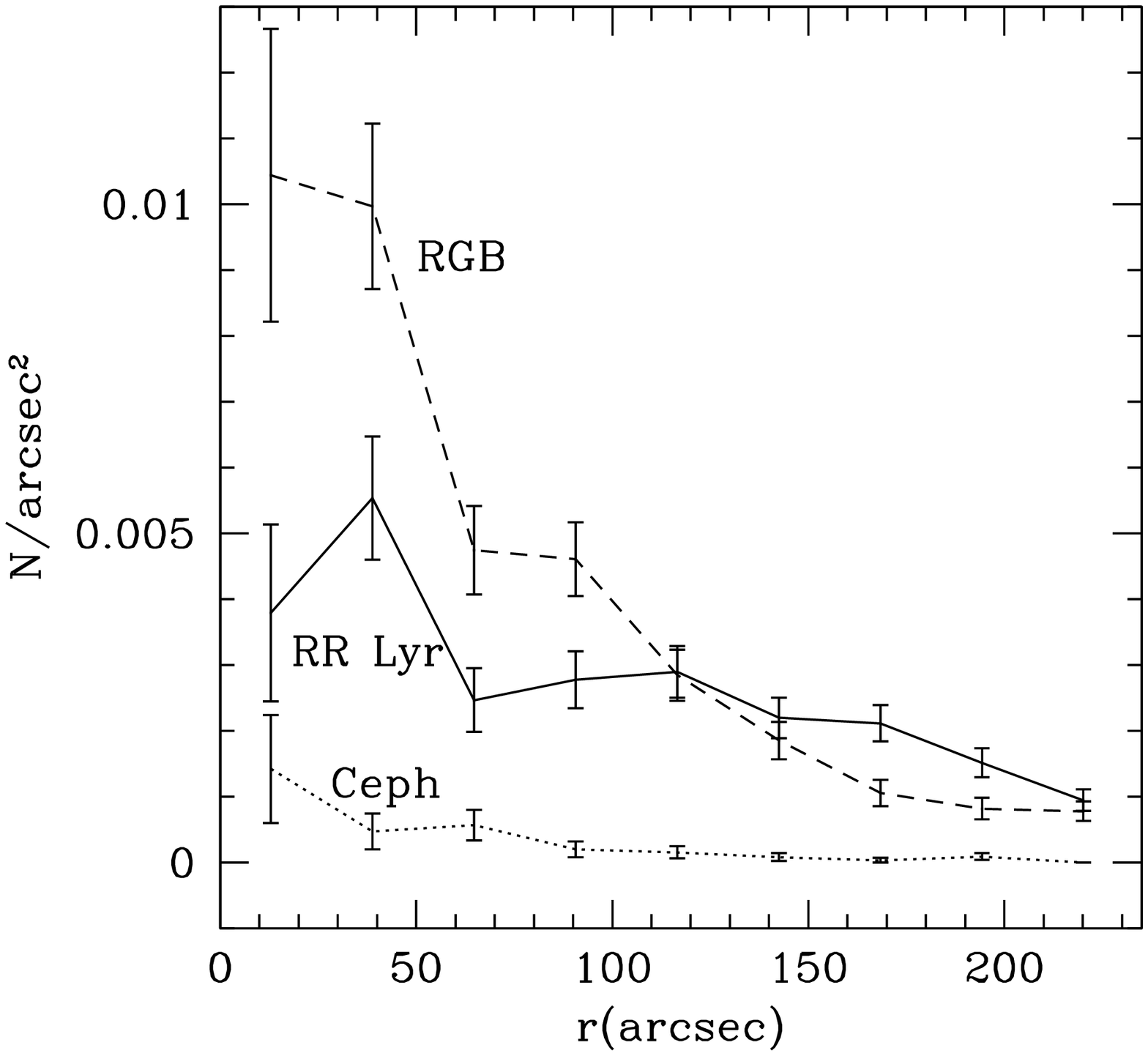}
\caption{Radial counts of RGB stars (I$\le$20.5, dashed line), RR Lyrae 
variable star candidates and Cepheid variables.}
\label{perfil}
\end{figure}

\clearpage

\begin{figure}
\plotone{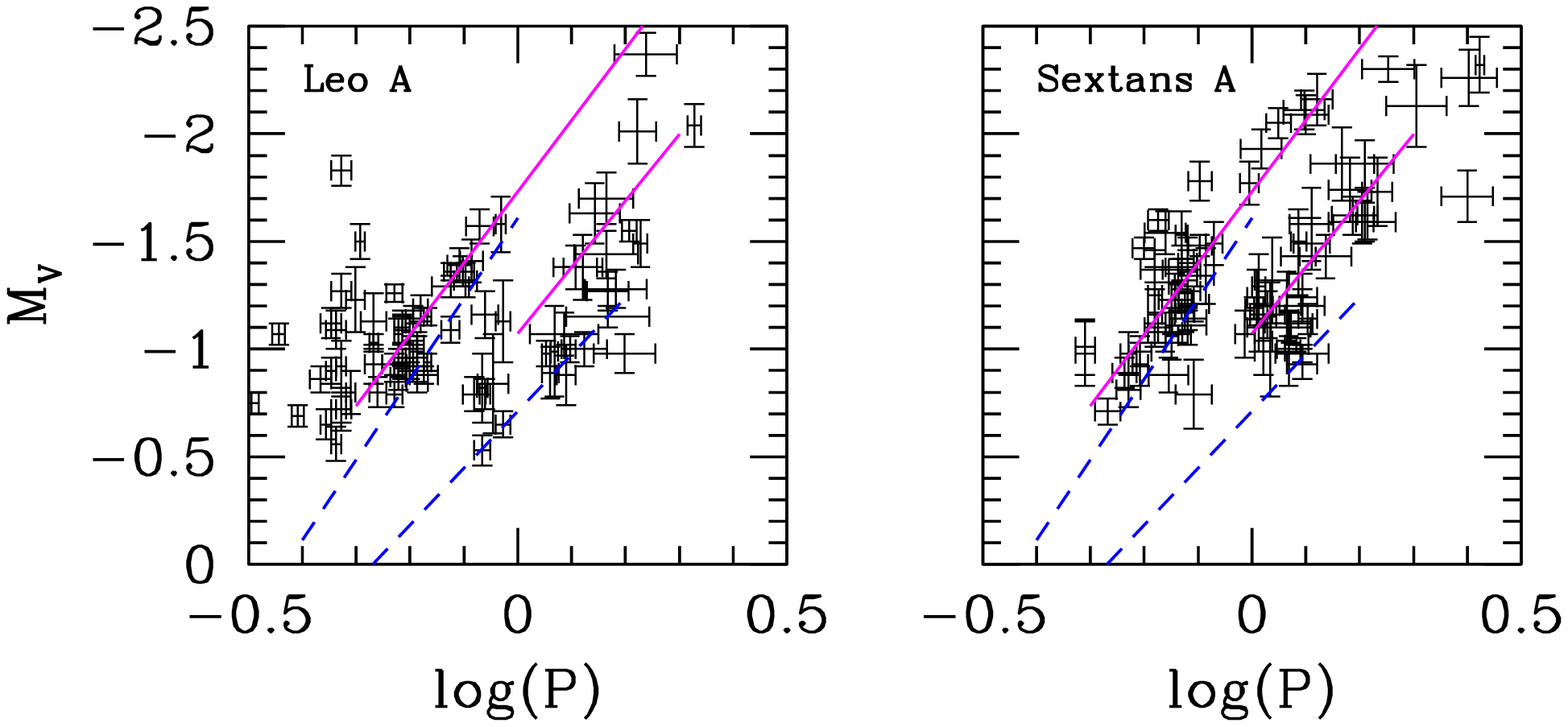}
\caption{Period-luminosity (PL) diagrams for the Leo A (left) and 
Sextans~A (right) Cepheids. Distance modulus $(m-M)_V=$ 24.54 and
25.72, according to Dolphin et al.\ (2002) and Dolphin et al.\ (2003)
respectively, have been assumed.  The PL relations obtained for dSph
AC and OGLE SMC s-pCC are represented as in Figure~\ref{pl}.}
\label{leoa_sex}
\end{figure}

\clearpage

\begin{deluxetable}{lrrrr}
\tablewidth{0pt}
\tablecaption{Observation Log}
\tablehead{
\colhead{Image ID} & \colhead{JD $-$2400000.0}& \colhead{${\rm T_{exp}}$ (sec)}& 
\colhead{Filter} & \colhead{seeing}}
\startdata

\cutinhead{{\it ESO Archival data}: 3.6m NTT w/EMMI, December 23, 1992}

emmi1992v1.imh & 48979.652773 & 720.000 & $V$ & 0.8\\ emmi1992v2.imh &
48979.663096 & 720.000 & $V$ & 0.9\\ emmi1992v3.imh & 48979.684380 &
720.000 & $V$ & 0.9\\

\cutinhead{LCO 100'', February 8-16, 1997}

clco2031 &   50489.537980 & 300.070 & $V$ & 1.1\\ 
clco2032 &   50489.545827 & 300.050 & $V$ & 1.0\\ 
clco2033 &   50489.550514 & 300.060 & $V$ & 0.9\\ 
clco2037 &   50489.569090 & 300.050 & $V$ & 1.0\\ 
clco2038 &   50489.573673 & 300.070 & $V$ & 1.1\\ 
clco2039 &   50489.578164 & 300.070 & $V$ & 1.2\\ 
clco3036 &   50490.529709 & 300.060 & $V$ & 1.1\\ 
clco3037 &   50490.534142 & 300.060 & $V$ & 1.1\\ 
clco3038 &   50490.538575 & 300.040 & $V$ & 0.9\\ 
clco3043 &   50490.573770 & 600.050 & $V$ & 1.1\\ 
clco5025 &   50492.584464 & 600.040 & $V$ & 1.1\\ 
clco6029 &   50493.545780 & 600.050 & $V$ & 1.4\\ 
clco6030 &   50493.553685 & 600.070 & $V$ & 1.4\\ 
clco7025 &   50494.532872 & 600.050 & $V$ & 1.6\\ 
clco8029 &   50495.563285 & 600.040 & $V$ & 1.9\\ 
clco9028 &   50496.579857 & 600.040 & $V$ & 1.6\\ 

clco1027 &   50488.564815 & 300.050 & $I$ & 0.9\\ 
clco2034 &   50489.555583 & 300.050 & $I$ & 0.8\\ 
clco2035 &   50489.560016 & 300.050 & $I$ & 0.9\\ 
clco2036 &   50489.564449 & 300.060 & $I$ & 0.8\\ 
clco3040 &   50490.549651 & 600.050 & $I$ & 1.0\\ 
clco3041 &   50490.557555 & 600.050 & $I$ & 1.0\\ 
clco3042 &   50490.565472 & 600.050 & $I$ & 0.9\\ 
clco5024 &   50492.576397 & 600.050 & $I$ & 1.2\\ 
clco6027 &   50493.528779 & 600.060 & $I$ & 1.2\\ 
clco6028 &   50493.537598 & 600.060 & $I$ & 1.0\\ 
clco7024 &   50494.524724 & 600.060 & $I$ & 1.3\\ 
clco8030 &   50495.571467 & 600.050 & $I$ & 1.6\\ 
clco9027 &   50496.571801 & 600.060 & $I$ & 1.2\\ 

\cutinhead{LCO 100'', TEK5, November 22-26, 1997}
blco2019 &   50774.518148 & 900.060 & $V$ & 1.3\\ 
blco2020 &   50774.529768 & 900.070 & $V$ & 0.9\\ 
blco3025 &   50775.639881 & 900.040 & $V$ & 1.0\\ 
blco3026 &   50775.651374 & 900.070 & $V$ & 0.9\\ 
blco4010 &   50776.640602 & 900.040 & $V$ & 0.7\\ 
blco4011 &   50776.652117 & 900.060 & $V$ & 0.8\\ 
blco5041 &   50777.603592 & 900.040 & $V$ & 0.8\\ 
blco5042 &   50777.615177 & 900.060 & $V$ & 0.8\\ 
blco6051 &   50778.574973 & 900.060 & $V$ & 1.3\\ 
blco6052 &   50778.586477 & 900.070 & $V$ & 1.1\\ 

\cutinhead{{\it ESO Archival data}: 3.6m NTT w/SUSI2, July 16, 1998}

susijul98v &   51011.931727 & 1800.000 & $V$ & 1.3\\ 
susijul98i &   51011.947320 & 801.193 & $I$ & 1.3\\ 

\cutinhead{{\it ESO Archival data}: 3.6m NTT w/SUSI2, October 28, 1998}

susinov98b1 &   51114.759908 & 399.999 & $B$ & 0.8\\ 
susinov98b2 &   51114.765005 & 400.000 & $B$ & 0.7\\ 
susinov98b3 &   51114.770044 & 399.999 & $B$ & 0.7\\ 
susinov98b4 &   51114.775088 & 400.000 & $B$ & 0.7\\ 
susinov98b5 &   51114.780112 & 399.999 & $B$ & 0.9\\ 
susinov98b6 &   51114.785211 & 400.000 & $B$ & 0.7\\ 

susinov98v1 &   51114.789725 & 300.000 & $V$ & 0.9\\ 
susinov98v2 &   51114.793617 & 299.999 & $V$ & 0.7\\ 

susinov98i1 &   51114.799479 & 399.999 & $I$ & 0.7\\ 
susinov98i2 &   51114.804636 & 399.999 & $I$ & 0.7\\ 
susinov98i3 &   51114.809681 & 399.999 & $I$ & 0.7\\
 
\cutinhead{LCO 100'', TEK5, September 11-14, 1998}
dlco381 &  2451070.85641682 & 900.050 & $B$ & 1.5 \\
dlco382 &  2451070.87038687 & 900.060 & $B$ & 1.5 \\

\cutinhead{{\it ESO Archival data}: VLT-Antu w/FORS1, August 20, 1999}

fors1999b1 &   51410.814718 & 599.976 & $B$ & 0.8\\ 
fors1999b2 &   51410.822354 & 600.000 & $B$ & 0.9\\ 
fors1999b3 &   51410.829979 & 600.000 & $B$ & 0.9\\ 

fors1999r1 &   51410.836018 & 299.989 & R & 0.8\\ 
fors1999r2 &   51410.840160 & 300.000 & R & 0.9\\ 

\cutinhead{VLT-Antu w/FORS1, September 15, 1999}

phob &   51436.731704 & 300.000 & $B$ & 1.0 \\ 
phoi &   51436.898504 & 300.000 & $I$ & 0.8\\ 
phor &   51436.739953 & 299.998 & R & 0.9\\ 
phov &   51436.735826 & 300.000 & $V$ & 0.6\\ 

\cutinhead{{\it ESO Archival data}: VLT-Antu w/FORS1, July 7-8, 2000}

fors2000b1 &   51550.540040 & 299.993 & $B$ & 0.8\\ 
fors2000b2 &   51550.544182 & 299.992 & $B$ & 0.7\\ 
fors2000b3 &   51551.567297 & 299.986 & $B$ & 0.7\\
fors2000v1 &   51550.529901 & 399.982 & $V$ & 0.8\\ 
fors2000v2 &   51550.535205 & 399.983 & $V$ & 0.8\\ 
fors2000v3 &   51551.562306 & 399.972 & $V$ & 0.8\\ 

\cutinhead{LCO 100'', TEK5, November 25-29, 2000}

b2ccd2038 &   51873.612653 & 1200.030 & $B$ & 0.9\\ 
b2ccd3041 &   51874.630813 & 1200.030 & $B$ & 0.8\\ 
b2ccd4040 &   51875.644344 & 1200.030 & $B$ & 0.8\\ 
b2ccd5042 &   51876.688254 & 1200.040 & $B$ & 0.9\\ 
b2ccd6037 &   51877.621985 & 1200.030 & $B$ & 0.8\\ 

b2ccd2034 &   51873.579240 &  450.040 & $V$ & 0.8\\ 
b2ccd2035 &   51873.585722 &  450.040 & $V$ & 0.8\\ 
b2ccd3037 &   51874.597146 &  450.040 & $V$ & 0.8\\ 
b2ccd3038 &   51874.603384 &  450.040 & $V$ & 0.8\\ 
b2ccd4037 &   51875.614959 &  900.030 & $V$ & 0.8\\ 
b2ccd5043 &   51876.701564 &  900.030 & $V$ & 0.8\\ 
b2ccd6038 &   51877.635306 &  900.030 & $V$ & 0.9\\ 

b2ccd2036f &   51873.593001 & 600.030 & $I$ & 0.7\\ 
b2ccd2037f &   51873.601068 & 600.040 & $I$ & 0.7\\ 
b2ccd3039f &   51874.611278 & 600.030 & $I$ & 0.7\\ 
b2ccd3040f &   51874.619240 & 600.040 & $I$ & 0.7\\ 
b2ccd4038f &   51875.624854 & 600.030 & $I$ & 0.7\\ 
b2ccd4039f &   51875.632817 & 600.050 & $I$ & 0.7\\ 
b2ccd5044f &   51876.711366 & 600.030 & $I$ & 1.0\\ 
b2ccd5045f &   51876.719283 & 600.040 & $I$ & 0.8\\ 
b2ccd6039f &   51877.645086 & 600.030 & $I$ & 0.8\\ 
b2ccd6040f &   51877.653037 & 600.040 & $I$ & 0.9\\

\cutinhead{LCO 100'', TEK5, December 13-21, 2001}

fccd1063 &   52256.648676 & 1200.040 & $V$ & 1.1\\ 
fccd2051 &   52257.637329 & 1200.030 & $V$ & 0.8\\ 
fccd3047 &   52258.643909 & 1200.040 & $V$ & 1.0\\ 
fccd4048 &   52259.614716 & 1200.040 & $V$ & 1.0\\ 
fccd5060 &   52260.647428 & 1200.050 & $V$ & 1.7\\ 
fccd6052 &   52261.662607 & 1200.030 & $V$ & 1.0\\ 
fccd6053 &   52261.680418 & 1200.040 & $V$ & 1.3\\ 
fccd7038 &   52262.556439 & 1200.030 & $V$ & 0.9\\ 
fccd7049 &   52262.710307 & 1200.030 & $V$ & 1.2\\ 
fccd9056 &   52264.678295 & 1200.030 & $V$ & 0.9\\ 
fccd10058 &   52265.655478 & 1200.030 & $V$ & 1.2\\ 
fccd11053 &   52266.649246 & 1200.030 & $V$ & 1.0\\

fccd1064 &   52256.664253 & 900.030 & $I$ & 0.9\\ 
fccd1065 &   52256.677401 & 900.030 & $I$ & 1.1\\ 
fccd2052 &   52257.651575 & 900.030 & $I$ & 0.8\\ 
fccd2053 &   52257.663323 & 900.040 & $I$ & 0.7\\ 
fccd3045 &   52258.618239 & 900.040 & $I$ & 0.8\\ 
fccd3046 &   52258.629755 & 900.030 & $I$ & 0.7\\ 
fccd4049 &   52259.628071 & 900.030 & $I$ & 0.9\\ 
fccd4050 &   52259.639599 & 900.040 & $I$ & 0.8\\ 
fccd5061 &   52260.662381 & 900.030 & $I$ & 1.2\\ 
fccd5062 &   52260.675436 & 900.040 & $I$ & 1.3\\ 
fccd6054 &   52261.694225 & 900.030 & $I$ & 1.5\\ 
fccd6055 &   52261.705672 & 900.050 & $I$ & 1.4\\ 
fccd7039 &   52262.569945 & 900.030 & $I$ & 0.7\\ 
fccd7040 &   52262.581727 & 900.040 & $I$ & 0.8\\ 
fccd7047 &   52262.684973 & 900.040 & $I$ & 1.2\\ 
fccd7048 &   52262.696615 & 900.040 & $I$ & 1.2\\ 
fccd9057 &   52264.691743 & 900.030 & $I$ & 0.9\\
fccd9058 &   52264.703143 & 900.030 & $I$ & 0.8\\ 
fccd10059 &   52265.669088 & 900.030 & $I$ & 1.1\\ 
fccd10060 &   52265.680442 & 900.030 & $I$ & 1.5\\ 
fccd11054 &   52266.663944 & 900.030 & $I$ & 0.8\\ 
fccd11055 &   52266.505285 & 900.030 & $I$ & 0.8\\ 

\enddata
\label{obslog}
\end{deluxetable}

\clearpage

\begin{deluxetable}{lrrrrrrr}
\tablecaption{Confirmed variable stars}
\tablehead{
\colhead{ID\tablenotemark{a}} & \colhead{X}& \colhead{Y}&  \colhead{Period}&\colhead{mean $B$}& \colhead{mean $V$}& \colhead{mean $I$}& \colhead{Sub-type}}
\startdata
\cutinhead{Cepheid (AC or s-pCC) variable stars}

  7382\tablenotemark{c} &  732.369 & 1231.179 & 1.69340 &   21.50  &  21.28 & 20.79 &   s-pCC\\
  7275 & 1222.072 & 1220.612 & 1.09550 &   21.69  &  21.42 & 20.96 &   s-pCC\\
  9027\tablenotemark{d} &  914.971 & 1376.685 & 1.55840 &   21.74  &  21.47 & 20.96 &   s-pCC\\
  8117 & 1166.702 & 1292.789 & 0.80340 &   21.61  &  21.50 & 21.21 & s-pCC  \\
  7452 &  925.259 & 1237.570 & 1.45070 &   21.97  &  21.62 & 21.10 &  s-pCC \\
  7922\tablenotemark{e} &  911.648 & 1276.491 & 1.29860 &   22.09  &  21.67 & 21.18 &  s-pCC \\
  7342 &  774.810 & 1227.815 & 1.32315 &   22.15  &  21.69 & 21.08 &  s-pCC \\
  7056\tablenotemark{f} &  942.529 & 1202.454 & 1.07640 &   22.18  &  22.17 & 21.78 &  s-pCC\tablenotemark{g} \\
  3649 & 1664.440 &  755.690 & 1.35170 &   22.52  &  22.27 & 21.70 & AC \\
 10800 & 1292.180 & 1572.292 & 1.15280\tablenotemark{b} &   22.95  &  22.64 & 22.10 &  AC \\
 10800 & 1292.180 & 1572.292 & 7.54886 &   22.89  &  22.66 & 22.11 &   \\
  6793 & 1129.687 & 1176.729 & 0.62118\tablenotemark{b} &   23.14  &  22.66 & 21.90 & AC \\
  6793 & 1129.687 & 1176.729 & 1.63969 &   23.18  &  22.65 & 21.89 &   \\
 12003 & 1557.913 & 1755.501 & 0.57465 &   23.01  &  22.71 & 22.25 & AC  \\
 10527 & 1624.325 & 1535.367 & 0.82080 &   23.03  &  22.79 & 22.25 & AC \\
  3441 & 1556.875 &  714.576 & 0.36438 &   23.32  &  22.89 & 22.26 & AC \\
103951 & 1238.072 & 1306.065 & 0.57113 &   23.04  &  22.90 & 22.56 &  AC \\
  4515 & 1165.694 &  907.197 & 0.83070 &   23.20  &  22.91 & 22.35 &  AC \\
 11219 &  918.179 & 1633.937 & 0.63985 &   23.16  &  23.05 & 22.68 &  AC  \\
  7993 &  906.453 & 1282.576 & 0.13553 &   23.37  &  23.09 & 22.53 &   \\
  7993 &  906.453 & 1282.576 & 0.73566\tablenotemark{b} &   23.33  &  23.08 & 22.56 & AC  \\
  7993 &  906.453 & 1282.576 & 0.73927\tablenotemark{b} &   23.34  &  23.08 & 22.55 & AC  \\
  5818 &  660.618 & 1082.305 & 0.67564\tablenotemark{b} &   23.50  &  23.17 & 22.59 & AC  \\
  5818 &  660.618 & 1082.305 & 1.97859 &   23.47  &  23.13 & 22.58 &   \\

\cutinhead{Sample RR Lyrae}

104248 & 1130.850 & 1365.450 &  0.59045 &  23.92  &  23.62 & 22.99 &   \\
  6247 & 1324.488 & 1126.068 &  0.71405 &  23.98  &  23.76 & 23.12 &   \\ 
  4439 & 1116.305 &  896.573 &  0.75800 &  23.77  &  23.56 & 22.97 &   \\
  7008 & 1895.248 & 1198.236 &  0.63165 &  23.97  &  23.61 & 22.97 &   \\

\cutinhead{Candidate Long Period Variables}

 10647 & 1969.314 & 1551.632 & & 19.80 & 19.69 & 19.05 & \\
 11200 & 1692.356 & 1631.371 & & 22.21 & 20.49 & 18.87 & \\
  8563 &  964.725 & 1334.615 & & 22.15 & 20.96 & 19.25 & \\ 
 11263 & 1389.933 & 1638.590 & & 25.57 & 21.19 & 18.90 & \\
 13288 & 1752.993 & 1980.251 & & 21.93 & 22.23 & 21.65 & \\ 
  3359 &  522.283 &  696.066 & & 23.17 & 22.54 & 21.92 & \\ 

\cutinhead{Candidate eclipsing binary}

  2347 &  936.704 &  429.077 & & 22.35 & 22.39 & 21.68 & \\ 
\enddata

\tablenotetext{a}{Stars listed multiple times have more than one possible
 period producing an acceptable light curve}
\tablenotetext{b}{Most likely period according to the period-luminosity diagram}
\tablenotetext{c}{Corresponding to MGA99 candidate \#2981 }
\tablenotetext{d}{Corresponding to MGA99 candidate \#3814 }
\tablenotetext{e}{Corresponding to MGA99 candidate \#3200 }
\tablenotetext{f}{Corresponding to MGA99 candidate \#2697 }
\tablenotetext{g}{$M_V$=-0.99, and therefore, according to the criterium 
of separation between s-pCC and AC at $M_V=-1$ stablished in
Section~\ref{nature_cep}, this variable should be classified as AC. Its
position in the PL diagram, however, is closer to the s-pCC locus, and
therefore, given that the separation at $M_V$=-0.99 is rather
arbitrary, we tentatively classify this star as s-pCC. This is the
only variable basically at the cutoff between AC and s-pCC.}
\label{candidates}
\end{deluxetable}

\clearpage

\begin{deluxetable}{rrr}
\tablewidth{0pt}
\tablecaption{B Photometry of Cepheid variable stars}
\tablehead{
\colhead{HJD-2400000} & \colhead{B} & \colhead{$\sigma(B)$}}
\startdata
\cutinhead{7382}
 51873.61265  & 21.662  & 0.014 \\
 51410.81472  & 21.092  & 0.005 \\
 51410.82235  & 21.131  & 0.006 \\
 51410.82998  & 21.123  & 0.006 \\
 51436.73170  & 21.616  & 0.012 \\
 51550.54004  & 21.762  & 0.016 \\
\enddata        	  	 
\label{bphotometry}
\tablecomments{Table presented in its enterity in the electronic edition of the AJ only}	  	 
\end{deluxetable}

\clearpage

\begin{deluxetable}{rrr}
\tablewidth{0pt}
\tablecaption{V Photometry of Cepheid variable stars}
\tablehead{
\colhead{HJD-2400000} & \colhead{V} & \colhead{$\sigma(V)$}}
\startdata
\cutinhead{7382}

 48979.65277 &  21.289 &  0.023 \\ 
 48979.66310 &  21.355 &  0.031 \\ 
 48979.68438 &  21.376 &  0.028 \\ 
 50489.53798 &  21.025 &  0.023 \\ 
 50489.54583 &  20.973 &  0.018 \\ 
 50489.55051 &  20.974 &  0.019 \\ 	 	
\enddata
\label{vphotometry}
\tablecomments{Table presented in its enterity in the electronic edition of the AJ only}
\end{deluxetable}

\clearpage

\begin{deluxetable}{rrr}
\tablewidth{0pt}
\tablecaption{I Photometry of Cepheid variable stars}
\tablehead{
\colhead{HJD-2400000} & \colhead{I} & \colhead{$\sigma(I)$}}
\startdata

\cutinhead{7382}

 50488.56482 &  20.781 &  0.061 \\ 
 50489.55558 &  20.559 &  0.045 \\ 
 50489.56002 &  20.534 &  0.041 \\ 
 50489.56445 &  20.570 &  0.041 \\ 
 50490.54965 &  20.849 &  0.024 \\ 
 50490.55756 &  20.847 &  0.024 \\ 
\enddata
\label{iphotometry}
\tablecomments{Table presented in its enterity in the electronic edition of the AJ only}
\end{deluxetable}

\end{document}